\documentclass[journal,10pt,twoside]{IEEEtran} 
\pdfoutput=1

\usepackage{amsmath,afterpage}
\usepackage{amsfonts}
\usepackage{amssymb}
\usepackage{amsthm}
\usepackage[usenames,dvipsnames,table]{xcolor}
\usepackage{graphicx, color, transparent}
\usepackage{tabularx}
\usepackage{makecell}
\usepackage{multirow}
\usepackage{pbox}
\usepackage{tikz}
\usetikzlibrary{plotmarks}
\usepackage{pgfplots}
\usetikzlibrary{calc}
\usetikzlibrary{shapes,arrows}
\usetikzlibrary{decorations.markings}
\pgfplotsset{compat=1.10}
\usepackage{amsthm}
\usepackage{cite}
\makeatletter
\newtheorem*{rep@theorem}{\rep@title}
\newcommand{\newreptheorem}[2]{
\newenvironment{rep#1}[1]{
 \def\rep@title{\Cref{##1}}
 \begin{rep@theorem}}
 {\end{rep@theorem}}}
\newcommand*{\textlabel}[2]{
  \edef\@currentlabel{#1}
  \phantomsection
  #1\label{#2}
}
\makeatother

\makeatletter
\newcommand{\wast}{\bBigg@{2}}
\newcommand{\Wast}{\bBigg@{3}}
\newcommand{\vast}{\bBigg@{4}}
\newcommand{\Vast}{\bBigg@{5}}
\makeatother
\usepackage{todonotes}

\newcommand{\Enc}{\mathsf{Enc}}
\newcommand{\Dec}{\mathsf{Dec}}
\newcommand{\Bin}{\mathsf{Bin}}
\newcommand{\overbar}[1]{\mkern 1.5mu\overline{\mkern-1.5mu#1\mkern-1.5mu}\mkern 1.5mu}

\usepackage{enumerate}
\usepackage{url}
\usepackage{cleveref}
\crefname{table}{table}{tables}
\Crefname{table}{Table}{Tables}
\crefname{figure}{figure}{figures}
\Crefname{figure}{Figure}{Figures}
\crefname{section}{section}{sections}
\Crefname{section}{Section}{Sections}

\newcommand*\xor{\mathbin{\oplus}}

\newtheorem{lemma}{Lemma}

\newtheorem{remark}{Remark}

\theoremstyle{definition}

\theoremstyle{plain}
\newtheorem{theorem}{Theorem}
\newreptheorem{theorem}{Theorem}
\newreptheorem{lemma}{Lemma}

\begin{document}
\title{Privacy, Secrecy, and Storage with Multiple\\ Noisy Measurements of Identifiers}

\author{Onur~G\"unl\"u,~\IEEEmembership{Student Member,~IEEE,} and
        Gerhard~Kramer,~\IEEEmembership{Fellow,~IEEE}
\thanks{Manuscript received August 30, 2017; revised January 26, 2018 and April 28, 2018; accepted April 28, 2018. O. G\"unl\"u was supported by the German Research Foundation (DFG) through the project HoliPUF under the grant KR3517/6-1. G. Kramer was supported by an Alexander von Humboldt Professorship endowed by the German Federal Ministry of Education and Research. Parts of this paper were presented at the 2015 IEEE Conference on Communications and Network Security in \cite{bizimCNS}. The associate editor coordinating the review of this manuscript and approving it for publication was Dr. Tanya Ignatenko (\textit{Corresponding Author: Onur G\"unl\"u}).}
\thanks{The authors are with the Chair of Communications Engineering, Technical University of Munich, 80290 Munich, Germany (e-mail: \{onur.gunlu, gerhard.kramer\}@tum.de).}
}

\markboth{IEEE Transactions on Information Forensics and Security}{G\"unl\"u and Kramer: Privacy, Secrecy, and Storage with Multiple Noisy Measurements of Identifiers}

\maketitle

\begin{abstract}
 The key-leakage-storage region is derived for a generalization of a classic two-terminal key agreement model. The additions to the model are that the encoder observes a hidden, or noisy, version of the identifier, and that the encoder and decoder can perform multiple measurements. To illustrate the behavior of the region, the theory is applied to binary identifiers and noise modeled via binary symmetric channels. In particular, the key-leakage-storage region is simplified by applying Mrs.~Gerber's lemma twice in different directions to a Markov chain. The growth in the region as the number of measurements increases is quantified. The amount by which the privacy-leakage rate reduces for a hidden identifier as compared to a noise-free (visible) identifier at the encoder is also given. If the encoder incorrectly models the source as visible, it is shown that substantial secrecy leakage may occur and the reliability of the reconstructed key might decrease. 
\end{abstract}


\IEEEpeerreviewmaketitle
\section{Introduction} \label{sec:intro}
\IEEEPARstart{B}{iometric} identifiers can be used to authenticate or identify a user, and to generate secret keys \cite{Identifier}. Similarly, physical identifiers such as fine variations of ring oscillator (RO) outputs produce device ``fingerprints'' that can authenticate a device and generate keys \cite{bizimpaper,IntrinsicIP,pufintheory}. For instance, physical unclonable functions (PUFs) are physical identifiers that are cheaper and safer alternatives to key storage in non-volatile memories \cite{GassendThesis,PappuThesis}. Replacing biometric or physical identifiers, e.g., if the fingerprint is stolen, is often not possible \cite{IgnaTrans} or might require reconfigurable identifier designs \cite{ReconfigurablePUF}. Replaced physical identifiers may have correlated outputs with previous identifiers due to surrounding logic \cite{Systematicvariation}. One should, therefore, limit the information leaked about the identifier outputs, as well as the information leaked about the secret key.  

Consider the key agreement model introduced in \cite{AhlswedeCsiz} and \cite{Maurer} where two terminals observe dependent random variables and have access to a public communication link; an eavesdropper observes the messages, called \textit{helper data}, transmitted over this link. We consider a \textit{generated-secret} (GS) model and a \textit{chosen-secret} (CS) model. For the GS model, an encoder extracts a key from the source, while for the CS model a key that is independent of the source is given to the first terminal. The information through helper data about the secret key, called the {\textit{secrecy leakage}, should be negligible. The information leaked about the identifier, called the \textit{privacy leakage}, should be minimized so that an eavesdropper cannot obtain information about a secret key stored by another encoder that uses the same or a correlated identifier.   

The secret-key vs. privacy-leakage, or key-leakage, regions for the two models are given in \cite{IgnaTrans} and \cite{LaiTrans}. In addition to the secret-key and privacy-leakage rates, it is important to consider the amount of storage in the public link that is required for the decoder to reliably reconstruct the secret key \cite{csiszarnarayan}. The storage rate is generally equal to the privacy-leakage rate when we consider the GS model. Similarly, for the CS model, the storage rate is generally equal to the sum of the secret-key and privacy-leakage rates. The storage rate is different from the privacy-leakage rate for general (non-negligible) secrecy-leakage levels \cite{storage}, unlike for the negligible secrecy-leakage rate constraint considered in \cite{IgnaTrans} and \cite{LaiTrans}. We show that the storage and privacy-leakage rates are different also when the identifier is a \emph{remote} or \emph{hidden} source \cite[p.~118]{CsiszarKornerbook2011}, \cite[p.~78]{Bergerbook}.

Secret-key based user or device authentication with a privacy-leakage constraint is considered in \cite{KittipongTransIdent}. There is an assumption in \cite{KittipongTransIdent} that the eavesdropper has side information correlated with the identifier outputs, which is reasonable for biometric identifiers because they are continuously available for attacks. However, physical identifiers like PUFs are used for on-demand key reconstruction. Invasive attacks on PUFs also permanently change the identifier output \cite{PappuThesis}, so we assume that the eavesdropper cannot obtain information correlated with the PUF output. Key agreement with correlated side information at the eavesdropper has been studied in \cite{SecrecyviaSourcesandChannels,Blochpaper,Khisti}.

\subsection{Motivation}
Multiple measurements of biometric or physical identifiers at the decoder can substantially decrease the privacy-leakage and storage rates because less side information is required to reconstruct the secret key as compared to a single measurement. One obtains a diversity gain, corresponding to a gain in reliability, to combat erroneous measurements by averaging over different channels. One can also exploit the additional degrees of freedom by increasing the extracted secret-key size. The latter gain can be viewed as a multiplexing gain, in analogy to multiple antenna systems for wireless communications. Such gains in the achievable key-leakage rates are illustrated in \cite{bizimCNS} when there are multiple noisy measurements of the source at the decoder.

The above models assume that the encoder measures the ``true'' source. We propose that the true source, i.e., the ground truth, is instead hidden from the encoder and the encoder measures a noisy version of the source (see also discussions in \cite{bizimpaper} on key-binding with a hidden identifier, \cite{groundtruthauthentication} where a hidden source is considered for authentication, and \cite[Sec.~II]{permuter} for indirect rate-distortion problems with action-dependent side information). For example, many secrecy systems require multiple measurements at the encoder to obtain the ``noise-free'' output. As a second example, different systems may generate different sequences from the same identifier. 

Consider multiple encoders with independent channels from the hidden source to the corresponding encoder measurements. This is a valid scenario for biometric and physical identifiers due to differences in the environmental conditions when extracting secret keys by different encoders. An eavesdropper who wants to seize a secret key can use the information available from other encoders about the hidden source, which leads to privacy-leakage with respect to the hidden source rather than the noisy encoder measurements. 

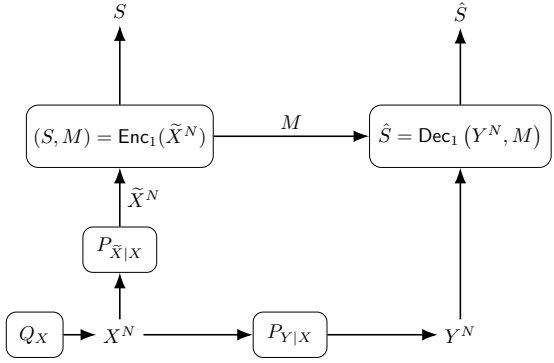
\begin{figure}
	\resizebox{0.84\linewidth}{!}{
		\centering
		\begin{tikzpicture}
		\node (so) at (-1.5,-3.5) [draw,rounded corners = 5pt, minimum width=1.0cm,minimum height=0.8cm, align=left] {$Q_X$};
		\node (a) at (0,0) [draw,rounded corners = 6pt, minimum width=3.2cm,minimum height=1.1cm, align=left] {$
			(S,M) = \Enc_1(\widetilde{X}^N)$};
		\node (c) at (3,-3.5) [draw,rounded corners = 5pt, minimum width=1.3cm,minimum height=0.8cm, align=left] {$P_{Y|X}$};
		\node (f) at (0,-2) [draw,rounded corners = 5pt, minimum width=1.3cm,minimum height=0.8cm, align=left] {$P_{\widetilde{X}|X}$};
		\node (b) at (6,0) [draw,rounded corners = 6pt, minimum width=3.2cm,minimum height=1.1cm, align=left] {$\hat{S} = \Dec_1\left(Y^N,M\right)$};
		\draw[decoration={markings,mark=at position 1 with {\arrow[scale=1.5]{latex}}},
		postaction={decorate}, thick, shorten >=1.4pt] (a.east) -- (b.west) node [midway, above] {$M$};
		\node (a1) [below of = a, node distance = 3.5cm] {$X^N$};
		\node (b1) [below of = b, node distance = 3.5cm] {$Y^N$};
		\draw[decoration={markings,mark=at position 1 with {\arrow[scale=1.5]{latex}}},
		postaction={decorate}, thick, shorten >=1.4pt] (so.east) -- (a1.west);
		\draw[decoration={markings,mark=at position 1 with {\arrow[scale=1.5]{latex}}},
		postaction={decorate}, thick, shorten >=1.4pt] (a1.north) -- (f.south);
		\draw[decoration={markings,mark=at position 1 with {\arrow[scale=1.5]{latex}}},
		postaction={decorate}, thick, shorten >=1.4pt] (f.north) -- (a.south) node [midway, right] {$\widetilde{X}^N$};
		\draw[decoration={markings,mark=at position 1 with {\arrow[scale=1.5]{latex}}},
		postaction={decorate}, thick, shorten >=1.4pt] (a1.east) -- (c.west);
		\draw[decoration={markings,mark=at position 1 with {\arrow[scale=1.5]{latex}}},
		postaction={decorate}, thick, shorten >=1.4pt] (c.east) -- (b1.west);
		\draw[decoration={markings,mark=at position 1 with {\arrow[scale=1.5]{latex}}},
		postaction={decorate}, thick, shorten >=1.4pt] (b1.north) -- (b.south);
		\node (a2) [above of = a, node distance = 2.2cm] {$S$};
		\node (b2) [above of = b, node distance = 2.2cm] {$\hat{S}$};
		\draw[decoration={markings,mark=at position 1 with {\arrow[scale=1.5]{latex}}},
		postaction={decorate}, thick, shorten >=1.4pt] (b.north) -- (b2.south);
		\draw[decoration={markings,mark=at position 1 with {\arrow[scale=1.5]{latex}}},
		postaction={decorate}, thick, shorten >=1.4pt] (a.north) -- (a2.south);
		\end{tikzpicture}
	}
	\caption{The GS model where a secret key is generated from a noisy identifier measurement.}\label{fig:biosystemgenerated}
\end{figure}

\subsection{Models for Identifier Outputs}
We study the physical and biometric identifier outputs that are independent and identically distributed (i.i.d.) according to a probability distribution with a discrete alphabet. These models are reasonable if one uses transform-coding algorithms, as in \cite{bizimtemperature}, to extract almost i.i.d. bits from PUFs under varying environmental conditions. Similar transform-coding based algorithms have been applied to biometric identifiers to obtain independent output symbols \cite{Transformbio}.

\subsection{Summary of Contributions and Organization} 
We extend the model of \cite{IgnaTrans} and \cite{LaiTrans} to include multiple noisy identifier measurements at the encoder and decoder. A summary of the main contributions is as follows.
\begin{itemize}
	\item We derive the key-leakage-storage regions for the GS and CS models with a hidden source; see Figs.~\ref{fig:biosystemgenerated} and \ref{fig:biosystemchosen} for the corresponding models. Our rate regions recover several results in the literature, including various results for a visible source without eavesdropper side information in \cite{AhlswedeCsiz,Maurer,csiszarnarayan,IgnaTrans,LaiTrans}. We further recover our previous results from \cite{bizimCNS} that studied the visible source model, as discussed in Section~\ref{sec:multimeasurecap}.
	\item We evaluate the rate region for a binary hidden source with multiple measurements at the decoder and a single noisy measurement at the encoder by applying Mrs. Gerber's lemma (MGL) \cite{WZ}. The analysis differs from \cite{IgnaTrans} and \cite{LaiTrans} because we need to apply MGL twice in different directions to a Markov chain rather than once. For measurement channels with a certain symmetry, we find the optimal auxiliary random variable for coding. 
	\item We show that a significant amount of secrecy might be leaked, and the reliability of the reconstructed key might decrease, if the visible source model is mistakenly used for multiple decoder measurements of a hidden source. Such a mistake leads to violations of the security and reliability constraints.
	\item Gains from having multiple measurements at the encoder are also illustrated. We show that gains in the secret-key rate can come at a large cost of storage.  
\end{itemize} 

This paper is organized as follows. In Section~\ref{sec:multimodel}, we describe our problem and develop the key-leakage-storage regions for the GS and CS models. The key-leakage-storage region of a binary hidden source with multiple measurements at the decoder is derived in Section~\ref{sec:multimeasurecap}. In Section~\ref{sec:examples}, we illustrate gains when using the hidden source model as compared to the visible one and depict the maximum secret-key rates achieved by having multiple encoder and decoder measurements. Achievability proofs and converses for the derived rate regions are given in Sections~\ref{sec:achProofs} and \ref{sec:convProofs}, respectively. 

\begin{figure}
	\resizebox{0.84\linewidth}{!}{
		\centering
		\begin{tikzpicture}
		\node (so) at (-1.5,-3.5) [draw,rounded corners = 5pt, minimum width=1.0cm,minimum height=0.8cm, align=left] {$Q_X$};
		\node (a) at (0,0) [draw,rounded corners = 6pt, minimum width=3.2cm,minimum height=1.1cm, align=left] {$
			M = \Enc_2(\widetilde{X}^N,S)$};
		\node (c) at (3,-3.5) [draw,rounded corners = 5pt, minimum width=1.3cm,minimum height=0.8cm, align=left] {$P_{Y|X}$};
		\node (f) at (0,-2) [draw,rounded corners = 5pt, minimum width=1.3cm,minimum height=0.8cm, align=left] {$P_{\widetilde{X}|X}$};
		\node (b) at (6,0) [draw,rounded corners = 6pt, minimum width=3.2cm,minimum height=1.1cm, align=left] {$\hat{S} = \Dec_2\left(Y^N,M\right)$};
		\draw[decoration={markings,mark=at position 1 with {\arrow[scale=1.5]{latex}}},
		postaction={decorate}, thick, shorten >=1.4pt] (a.east) -- (b.west) node [midway, above] {$M$};
		
		\node (a1) [below of = a, node distance = 3.5cm] {$X^N$};
		\node (b1) [below of = b, node distance = 3.5cm] {$Y^N$};
		\draw[decoration={markings,mark=at position 1 with {\arrow[scale=1.5]{latex}}},
		postaction={decorate}, thick, shorten >=1.4pt] (so.east) -- (a1.west);
		\draw[decoration={markings,mark=at position 1 with {\arrow[scale=1.5]{latex}}},
		postaction={decorate}, thick, shorten >=1.4pt] (a1.north) -- (f.south);
		\draw[decoration={markings,mark=at position 1 with {\arrow[scale=1.5]{latex}}},
		postaction={decorate}, thick, shorten >=1.4pt] (f.north) -- (a.south) node [midway, right] {$\widetilde{X}^N$};
		\draw[decoration={markings,mark=at position 1 with {\arrow[scale=1.5]{latex}}},
		postaction={decorate}, thick, shorten >=1.4pt] (a1.east) -- (c.west);
		\draw[decoration={markings,mark=at position 1 with {\arrow[scale=1.5]{latex}}},
		postaction={decorate}, thick, shorten >=1.4pt] (c.east) -- (b1.west);
		\draw[decoration={markings,mark=at position 1 with {\arrow[scale=1.5]{latex}}},
		postaction={decorate}, thick, shorten >=1.4pt] (b1.north) -- (b.south);
		\node (a2) [above of = a, node distance = 2.2cm] {$S$};
		\node (b2) [above of = b, node distance = 2.2cm] {$\hat{S}$};
		\draw[decoration={markings,mark=at position 1 with {\arrow[scale=1.5]{latex}}},
		postaction={decorate}, thick, shorten >=1.4pt] (b.north) -- (b2.south);
		\draw[decoration={markings,mark=at position 1 with {\arrow[scale=1.5]{latex}}},
		postaction={decorate}, thick, shorten >=1.4pt] (a2.south) -- (a.north);
		\end{tikzpicture}
	}
	\caption{The CS model where a secret key is given to the encoder together with a noisy identifier measurement.}\label{fig:biosystemchosen}
\end{figure}
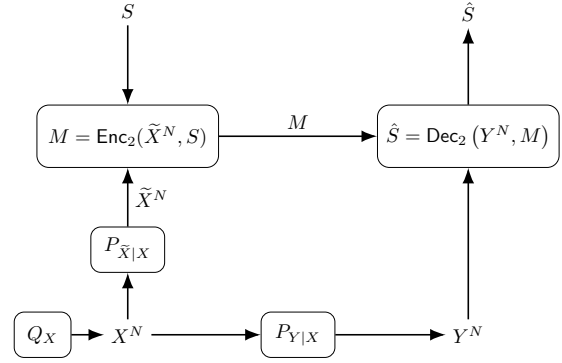

\subsection{Notation}
Upper case letters represent random variables and lower case letters their realizations. Superscripts denote a string of variables, e.g., $\displaystyle X^N\!=\!X_1\ldots X_i\ldots X_N$, and subscripts denote the position of a variable in a string. A random variable $\displaystyle X$ has probability distribution $\displaystyle Q_X$ or $\displaystyle P_X$. Calligraphic letters such as $\displaystyle \mathcal{X}$ denote sets; set sizes are written as $\displaystyle |\mathcal{X}|$ and set complements are denoted as $\displaystyle {\mathcal{X}}^{c}$. $\displaystyle \mathcal{T}_{\epsilon}^{N}(Q_X)$ denotes the set of length-$N$ letter-typical sequences with respect to the probability distribution $\displaystyle Q_X$ and the positive number $\displaystyle \epsilon$ \cite[Ch. 3]{masseylecturenotes}, \cite{orlitsky}. $H_b(x)\!=\!-x\log x\!-\! (1\!-\!x)\log (1\!-\!x)$ is the binary entropy function and $\displaystyle H_b^{-1}(\cdot)$ denotes its inverse with range $[0,0.5]$. The $*$-operator is defined as $\displaystyle p\!*\!x\! =\! p(1\!-\!x)\!+\!(1\!-\!p)x$. $\text{Unif}\,[1\!:\!N]$ denotes the uniform distribution over the integers $\displaystyle 1,2,\ldots,N$. The $\xor$-operator denotes modulo-2 summation. 

\section{System Models and Rate Regions} \label{sec:multimodel}

\subsection{System Models}
Consider a discrete memoryless source that generates i.i.d. symbols $\displaystyle X^{N}$ from a finite set $\displaystyle \mathcal{X}$ according to a probability distribution $\displaystyle Q_X$. Identifier outputs are noisy due to, for instance, cuts on a finger. The noise at the encoder and decoder is modeled as memoryless channels $\displaystyle P_{\widetilde{X}|X}$ and $\displaystyle P_{Y|X}$, respectively. The outputs of $\displaystyle P_{\widetilde{X}|X}$ and $\displaystyle P_{Y|X}$ are, respectively, the strings $\displaystyle \widetilde{X}^{N}$ with realizations from a finite set $\displaystyle \mathcal{\widetilde{X}}^N$, and $\displaystyle Y^{N}$ with realizations from a finite set $\displaystyle \mathcal{Y}^N$. We thus have
\vspace*{-0.1cm}
\begin{align}
	P_{\widetilde{X}^{N}X^{N}Y^{N}}&(\tilde{x}^{N},x^{N},y^{N})\nonumber\\
	&=\prod_{i=1}^{N}P_{\widetilde{X}|X}(\tilde{x}_i|x_i)Q_X(x_i)P_{Y|X}(y_i|x_i)\text{.}
\end{align}
The distributions $\displaystyle P_{\widetilde{X}|X}$ and $\displaystyle P_{Y|X}$ are assumed to be known for now, although we later study what happens if the encoder treats $\widetilde{X}^N$ as the true source.

In the GS model depicted in Fig.~\ref{fig:biosystemgenerated}, an encoder sees $\displaystyle \widetilde{X}^{N}$ and generates a secret key $S$ and helper data $M$ as $\displaystyle (S,M)\,{=}\,{\Enc_1}(\widetilde{X}^N)$, where ${\Enc_1}(\cdot)$ is an encoder mapping. The decoder estimates the key as $\displaystyle \hat{S}\,{=}\,{\Dec_1}(Y^N\!,M)$, where $\displaystyle \Dec_1(\cdot)$ is a decoder mapping. In the CS model shown in Fig.~\ref{fig:biosystemchosen}, $S$ is independent of $(X^N\text{, }\widetilde{X}^N\text{, }Y^N)$ and an encoder mapping ${\Enc_2}(\cdot)$ generates the helper data as $\displaystyle M\,{=}\, {\Enc}_{2}(\widetilde{X}^N,S)$. The decoder estimates the key as $\displaystyle \hat{S}\,{=}\, {\Dec_2}(Y^N,M)$, where $\Dec_2(\cdot)$ is a decoder mapping. 

A (secret-key, privacy-leakage, storage) rate triple $\displaystyle (R_s,R_l,R_m)$ is achievable if, given any $\delta\!>\!0$, there is some $N\!\geq\!1$, an encoder, and a decoder for which $\displaystyle R_s=\frac{\log|\mathcal{S}|}{N}$ and
\begin{alignat}{2}
&\Pr[S\ne\hat{S}] \leq \delta &&\qquad\quad (reliability) \label{eq:reliabilityconst}\\
&\frac{1}{N}I\left(S;M\right) \leq \delta &&\qquad\quad(weak\; secrecy)  \label{eq:secrecyconst}\\
&\frac{1}{N}I\left(X^N;M\right) \leq R_l+\delta \quad\quad\quad&&\qquad\quad (privacy)  \label{eq:privacyconst}\\
&\frac{1}{N} H(S) \geq R_s-\delta  \quad&&\qquad\quad (uniformity)\label{eq:uniformityconst}\\
&\frac{1}{N} H(M) \le R_m+\delta  &&\qquad\quad (storage).\label{eq:storageconst}
\end{alignat}

The key-leakage-storage region is the closure of the set of achievable rate tuples. We refer to models where $\widetilde{X}^N=X^N$ as visible source models (VSMs) and other cases as hidden source models (HSMs).

\subsection{Key-leakage-storage Regions}
We present the key-leakage-storage regions for the GS and CS models in Theorems~\ref{theo:regiongenerated} and ~\ref{theo:regionchosen}, respectively. The proofs of the theorems are given in Sections~\ref{sec:achProofs}-\ref{sec:convProofs}. We derive cardinality bounds for the auxiliary random variable in Appendix~\ref{app:cardinalitybound}. Using standard arguments, one can establish the convexity of the rate regions, i.e., there is no need for convexification via a time-sharing random variable.   

\begin{theorem}\label{theo:regiongenerated}
 The key-leakage-storage region for the GS model is 
\begin{subequations}
   \begin{align}
    \mathcal{R}_1\! =\! &\bigcup_{P_{U|\widetilde{X}}}\!\Big\{\left(R_s,R_l,R_m\right)\!\colon\! 0\leq R_s\leq I(U;Y),\nonumber\\
			 &R_l\geq I(U;X) - I(U;Y),\nonumber\\
			 &R_m\geq I(U;\widetilde{X})- I(U;Y)\Big\}\label{eq:rateregiongenerated}\\
		\text{where }&P_{U\widetilde{X}XY} = P_{U|\widetilde{X}}\cdot P_{\widetilde{X}|X}\cdot  Q_X\cdot P_{Y|X}\text{.}\label{eq:rateregiongenerated2}
    \end{align}
\end{subequations}
\end{theorem}

\begin{theorem}\label{theo:regionchosen}
 The key-leakage-storage region for the CS model is 
\begin{subequations} 
  \begin{align}
  \mathcal{R}_2\! =\! &\bigcup_{P_{U|\widetilde{X}}}\!\Big\{\left(R_s,R_l,R_m\right)\!\colon\! 0\leq R_s\leq I(U;Y),\nonumber\\
			 &R_l\geq I(U;X) - I(U;Y),\nonumber\\
			 &R_m\geq I(U;\widetilde{X})\Big\}\label{eq:rateregionchosen}\\
	\text{where }	 &P_{U\widetilde{X}XY} = P_{U|\widetilde{X}}\cdot P_{\widetilde{X}|X}\cdot Q_X\cdot P_{Y|X} \text{.} \label{eq:rateregionchosen2}
  \end{align}
\end{subequations}
\end{theorem}

\begin{remark}
 \normalfont The Markov conditions in (\ref{eq:rateregiongenerated2}) and (\ref{eq:rateregionchosen2}) state that $U\!-\!\widetilde{X}\!-\!X\!-\!Y$ forms a Markov chain. One may restrict the cardinality of the auxiliary random variable $U$ to $\displaystyle |\mathcal{U}|\!\leq\!|\mathcal{\widetilde{X}}|+2$ for both theorems. 
\end{remark}

\begin{remark}
	\normalfont The converses for Theorems~\ref{theo:regiongenerated} and \ref{theo:regionchosen} permit randomization at the encoder (see (\ref{eq:conversegsstoragerate})$(b)$ and (\ref{eq:conversecsstoragerate})$(b)$) and decoder (see (\ref{eq:fano})$(a)$). Since achievability requires no randomization, we may use deterministic encoders and decoders. The achievability of $\mathcal{R}_2$ follows directly from the achievability of $\mathcal{R}_1$ by using the key $S$ of the GS model as a key of a one-time pad to secure a chosen key and storing the output at rate $I(U;Y)$. 
\end{remark}

We recover the previous results in \cite{IgnaTrans} and \cite{LaiTrans} if $\widetilde{X}\!=\!X$ in both theorems so that the maximum achievable secret-key rate $\displaystyle I(\widetilde{X};Y)$ in these regions is at most $\displaystyle I(X;Y)$, which is the maximum achievable secret-key rate if the identifier $X^N$ is observed noise-free at the encoder. The minimum achievable privacy-leakage rate in these regions decreases as compared to in \cite{IgnaTrans} and \cite{LaiTrans} because $\displaystyle I(U;X)\leq I(U;\widetilde{X})$.

\section{Binary Identifier Measurements}\label{sec:multimeasurecap}
We evaluate the key-leakage-storage regions for a binary hidden source. The binary random sequence $\displaystyle \widetilde{X}^N$ corresponds to a single noisy measurement of the binary source $X^N$ at the encoder, and the random sequence $\displaystyle Y_{1:M_D}^N$ is the output of $M_D$ measurements of $X^N$ for $\displaystyle M_D\,{\geq}\,1$ at the decoder. We assume that the inverse channel $P_{X|\widetilde{X}}$ is a BSC, an assumption that is fulfilled if $P_X$ is uniform and $P_{\widetilde{X}|X}$ is a BSC. Moreover, we assume that the channel $\displaystyle P_{Y_{1:M_D}|X}$ can be decomposed into a mixture of BSCs  (i.e., binary-input symmetric memoryless channels \cite{chayat}, \cite{infcombining}), as described in \cite{bizimCNS} and illustrated below in Fig.~\ref{fig:corrBSCfig} for dependent BSCs. The former constraint lets us apply MGL to the Markov chain $U-\widetilde{X}-X$; the latter lets us apply an extension of MGL to the Markov chain $U-X-Y_{1:M_D}$. Recall that MGL is based on the result that, for any $0\leq p\leq 1$, the function
\begin{align}
	f(\nu)=H_b(p*H_b^{-1}(\nu))\label{eq:convexityboundxtildeu}
\end{align}	
is convex in $\nu$ for $0\leq\nu\leq 1$ \cite{WZ}.

Evaluating the key-leakage-storage regions corresponds to maximizing $\displaystyle I(U;Y_{1:M_D})$ and minimizing $\displaystyle I(U;\widetilde{X})$ for a fixed $\displaystyle I(U;X)$. It thus requires minimizing $H(Y_{1:M_D}|U)$ and maximizing $H(\widetilde{X}|U)$ for a fixed $H(X|U)$.

Let $\tilde{p}_i\!\in\![0,0.5]$ be the smaller transition probability from $U\!=\!u_i$ to $X=0$ or $X=1$ for $i\!\in\!\{1,2,\ldots,|\mathcal{U}|\}$. We have
\begin{align}
 &H(X|U) = \sum_{i=1}^{|\mathcal{U}|} P_U(u_i) H_b(\tilde{p}_i)\label{eq:hxgivenu}\\
 &H(Y_{1:M_D}|U) = \sum_{i=1}^{|\mathcal{U}|} P_U(u_i) g(\tilde{p}_i)\label{eq:hygivenu}
\end{align}
where
\begin{align}
g(\tilde{p}_i) = H(Y_{1:M_D}|U=u_i). \label{eq:definegw}
\end{align}

In the following, we first study dependent BSCs $\displaystyle P_{Y_{1:M_D}|X}$, which can be decomposed into a mixture of BSCs. We next discuss the convexity of the function $g(H_b^{-1}(\nu))$ in $\nu$ for binary-input channels $\displaystyle P_{Y_{1:M_D}|X}$ that can be decomposed into a mixture of BSCs to establish a tight lower bound on $H(Y_{1:M_D}|U)$ if we fix $H(X|U)$. Then, we simplify the key-leakage-storage regions of binary identifiers measured through such channels $\displaystyle P_{Y_{1:M_D}|X}$.

\subsection{Measurements Through Dependent BSCs}\label{subsec:newsubsecforexample}
We show that channels with multiple measurements of $X$ through dependent BSCs can be decomposed into a mixture of BSCs. For simplicity, consider $M_D\!=\!3$ with 
	\begin{align}
	\begin{bmatrix}
	Y_1\\
	Y_2\\
	Y_3\\
	\end{bmatrix}
	\!=\! X\! 
	\begin{bmatrix}
	1\\
	1\\
	1\\
	\end{bmatrix}
	\!\xor\! 
	\begin{bmatrix}
	B_1\\
	B_2\\
	B_3\\
	\end{bmatrix}\label{eq:corrBSCs}
	\end{align}
	where $B_1$, $B_2$, and $B_3$ are mutually dependent binary random variables that are jointly independent of $X$. We can decompose the channel (\ref{eq:corrBSCs}) into three BSCs, since we have 
	\begin{align}
	P_{Y_1Y_2Y_3|X}(y_1,y_2,y_3|x) = P_{Y_1Y_2Y_3|X}(\bar{y}_1,\bar{y}_2,\bar{y}_3|\bar{x})\label{eq:mixtureofBSCSfor3}
	\end{align}
	where $\bar{x}\!=\!1\!-\!x$ is the one's complement of $x$. Define
	\begin{align}
	&q_{y_1y_2y_3} = P_{Y_1Y_2Y_3|X}(y_1,y_2,y_3|0) \text{.}
	\end{align}
	The decomposed channel is depicted in Fig.~\ref{fig:corrBSCfig}, where the subchannel probabilities are
	\begin{align}
	&P_A(0) = q_{000} + q_{111}\\
	&P_A(1) = q_{001} + q_{110}\\
	&P_A(2) = q_{010} + q_{101}\\ 
	&P_A(3) = q_{011} + q_{100} 
	\end{align}
	and the crossover probabilities are $p_0\!=\!q_{111}/P_A(0)$, $p_1\!=\!q_{110}/P_A(1)$, $p_2\!=\!q_{101}/P_A(2)$, and $p_3\!=\!q_{100}/P_A(3)$.

	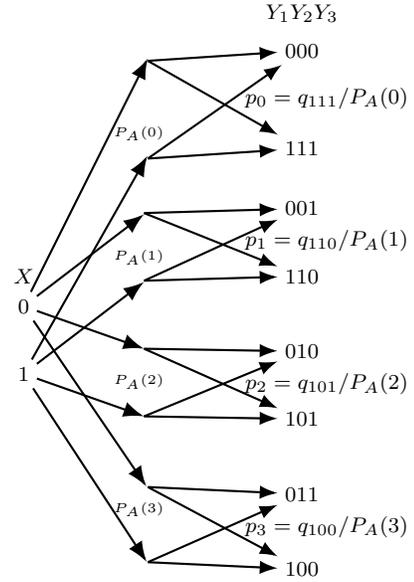
\begin{figure}[t]
		\centering
		\begin{tikzpicture}[auto,>=latex', scale = 1, transform shape]
		\footnotesize
		\node[align=center](X){$X$};
		\node[below of =X,node distance = 0.4cm](0){$0$};
		\node[below of =X,node distance = 1.3cm](1){$1$};
		\node[right of = X, node distance = 1.7cm](A110){};
		\node[above of = A110, node distance = 0.9cm](A001){};
		\node[above of = A110, node distance = 1.7cm](A111){};
		\node[above of = A110, node distance = 3cm](A000){};
		\node[below of = A110, node distance = 1cm](A010){};
		\node[below of = A110, node distance = 1.9cm](A101){};
		\node[below of = A110, node distance = 2.9cm](A011){};
		\node[below of = A110, node distance = 3.9cm](A100){};
		\node[right of = A001, node distance = 2cm](001){$001$};
		\node[right of = A110, node distance = 2cm](110){$110$};
		\node[right of = A010, node distance = 2cm](010){$010$};
		\node[right of = A101, node distance = 2cm](101){$101$};
		\node[right of = A000, node distance = 2cm](000){$000$};
		\node[right of = A111, node distance = 2cm](111){$111$};
		\node[right of = A011, node distance = 2cm](011){$011$};
		\node[right of = A100, node distance = 2cm](100){$100$};
		\node[above of = 001, node distance = 2.6cm](Y1Y2Y3){$Y_1Y_2Y_3$};
		\node[right of = X, node distance = 1.53cm](PAtemp){};
		\node[above of = PAtemp, node distance = 0.25cm](PA1){${\scriptscriptstyle P_A(1)}$};
		\node[below of = PA1, node distance = 1.65cm](PA2){${\scriptscriptstyle P_A(2)}$};
		\node[above of = PA1, node distance = 1.65cm](PA0){${\scriptscriptstyle P_A(0)}$};
		\node[below of = PA1, node distance = 3.35cm](PA3){${\scriptscriptstyle P_A(3)}$};
		\node[below of = 001, node distance = 0.43cm](p0temp){};
		\node[right of = p0temp, node distance = 0.33cm](p1){$p_1 = q_{110}/P_A(1)$};
		\node[below of = p1, node distance = 1.9cm](p2){$p_2 = q_{101}/P_A(2)$};
		\node[above of = p1, node distance = 1.9cm](p0){$p_0 = q_{111}/P_A(0)$};
		\node[below of = p1, node distance = 3.8cm](p3){$p_3 = q_{100}/P_A(3)$};
		
		\draw[decoration={markings,mark=at position 1 with {\arrow[scale=1.5]{latex}}},
		postaction={decorate}, thick, shorten >=1.4pt] (0) -- (A001); 
		\draw[decoration={markings,mark=at position 1 with {\arrow[scale=1.5]{latex}}},
		postaction={decorate}, thick, shorten >=1.4pt] (1) -- (A110); 
		\draw[decoration={markings,mark=at position 1 with {\arrow[scale=1.5]{latex}}},
		postaction={decorate}, thick, shorten >=1.4pt] (0) -- (A010); 
		\draw[decoration={markings,mark=at position 1 with {\arrow[scale=1.5]{latex}}},
		postaction={decorate}, thick, shorten >=1.4pt] (1) -- (A101);
		\draw[decoration={markings,mark=at position 1 with {\arrow[scale=1.5]{latex}}},
		postaction={decorate}, thick, shorten >=1.4pt] (0) -- (A000); 
		\draw[decoration={markings,mark=at position 1 with {\arrow[scale=1.5]{latex}}},
		postaction={decorate}, thick, shorten >=1.4pt] (1) -- (A111); 
		\draw[decoration={markings,mark=at position 1 with {\arrow[scale=1.5]{latex}}},
		postaction={decorate}, thick, shorten >=1.4pt] (0) -- (A011); 
		\draw[decoration={markings,mark=at position 1 with {\arrow[scale=1.5]{latex}}},
		postaction={decorate}, thick, shorten >=1.4pt] (1) -- (A100);
		\draw[decoration={markings,mark=at position 1 with {\arrow[scale=1.3]{latex}}},
		postaction={decorate}, thick, shorten >=1.4pt] ($(A001)+(-0.1,-0.06)$) -- (001);
		\draw[decoration={markings,mark=at position 1 with {\arrow[scale=1.3]{latex}}},
		postaction={decorate}, thick, shorten >=1.4pt] ($(A001)+(-0.1,-0.06)$) -- (110);
		\draw[decoration={markings,mark=at position 1 with {\arrow[scale=1.3]{latex}}},
		postaction={decorate}, thick, shorten >=1.4pt] ($(A110)+(-0.1,-0.06)$) -- (001);
		\draw[decoration={markings,mark=at position 1 with {\arrow[scale=1.3]{latex}}},
		postaction={decorate}, thick, shorten >=1.4pt] ($(A110)+(-0.1,-0.06)$) -- (110);
		\draw[decoration={markings,mark=at position 1 with {\arrow[scale=1.3]{latex}}},
		postaction={decorate}, thick, shorten >=1.4pt] ($(A000)+(-0.05,-0.135)$) -- (000);
		\draw[decoration={markings,mark=at position 1 with {\arrow[scale=1.3]{latex}}},
		postaction={decorate}, thick, shorten >=1.4pt] ($(A000)+(-0.05,-0.135)$) -- (111);
		\draw[decoration={markings,mark=at position 1 with {\arrow[scale=1.3]{latex}}},
		postaction={decorate}, thick, shorten >=1.4pt] ($(A111)+(-0.05,-0.135)$) -- (000);
		\draw[decoration={markings,mark=at position 1 with {\arrow[scale=1.3]{latex}}},
		postaction={decorate}, thick, shorten >=1.4pt] ($(A111)+(-0.05,-0.135)$) -- (111);
		\draw[decoration={markings,mark=at position 1 with {\arrow[scale=1.3]{latex}}},
		postaction={decorate}, thick, shorten >=1.4pt] ($(A010)+(-0.1,0.035)$) -- (010);
		\draw[decoration={markings,mark=at position 1 with {\arrow[scale=1.3]{latex}}},
		postaction={decorate}, thick, shorten >=1.4pt] ($(A010)+(-0.1,0.035)$) -- (101);
		\draw[decoration={markings,mark=at position 1 with {\arrow[scale=1.3]{latex}}},
		postaction={decorate}, thick, shorten >=1.4pt] ($(A101)+(-0.1,0.035)$) -- (010);
		\draw[decoration={markings,mark=at position 1 with {\arrow[scale=1.3]{latex}}},
		postaction={decorate}, thick, shorten >=1.4pt] ($(A101)+(-0.1,0.035)$) -- (101);
		\draw[decoration={markings,mark=at position 1 with {\arrow[scale=1.3]{latex}}},
		postaction={decorate}, thick, shorten >=1.4pt] ($(A011)+(-0.05,0.095)$) -- (011);
		\draw[decoration={markings,mark=at position 1 with {\arrow[scale=1.3]{latex}}},
		postaction={decorate}, thick, shorten >=1.4pt] ($(A011)+(-0.05,0.095)$) -- (100);
		\draw[decoration={markings,mark=at position 1 with {\arrow[scale=1.3]{latex}}},
		postaction={decorate}, thick, shorten >=1.4pt] ($(A100)+(-0.05,0.095)$) -- (011);
		\draw[decoration={markings,mark=at position 1 with {\arrow[scale=1.3]{latex}}},
		postaction={decorate}, thick, shorten >=1.4pt] ($(A100)+(-0.05,0.095)$) -- (100);
		
		\end{tikzpicture}
		\caption{$M_D=3$ dependent BSCs represented as a mixture of BSCs.}  
		\label{fig:corrBSCfig}
	\end{figure}

	More generally, we can decompose a channel with $M_D$ dependent BSC measurements into $\displaystyle 2^{M_D\!-\!1}$ subchannels each with output symbols such that one symbol is the one's complement of the other symbol. We define 
	\begin{align}
	q_{b^{M_D}} = P_{Y_{1:M_D}|X}(b^{M_D}|0)
	\end{align}
	for the length-$M_D$ binary string $\displaystyle b^{M_D} \!=\!b_0b_1\ldots b_{M_D\!-\!1}$ and
	\begin{align}
	P_A(a) = q_{{\Bin(a)}}\!+\!q_{{\overbar{\Bin(a)}}}
	\end{align}
	where $\displaystyle \overbar{\Bin(a)}$ is the one's complement of $\displaystyle \Bin(a)$ for $\displaystyle a\!=\!0,1,\ldots,2^{M_D\!-\!1}\!-\!1$. The crossover probability of the $a$-th subchannel is $p_a = q_{{\overbar{\Bin(a)}}}/P_A(a)$.

\subsection{Mixtures of BSCs}\label{subsec:channelmixtures}
Consider a channel $P_{Y_{1:M_D}|X}$ with a binary input and $M_D$ binary measurements as output, i.e., the channel has $2^{M_D}$ possible output symbols. We decompose the channel into $L=2^{M_D-1}$ BSCs as described above. We index these BSCs from $1$ to $L$. Let $A=a$ represent the BSC index chosen by the channel and let $p_a$ be the crossover probability of $a$-th subchannel. The conditional decoder-output entropy is
\begin{align}
 &H(Y_{1:M_D}|U) \stackrel{(a)}{=} H(Y_{1:M_D}A|U)\nonumber\\ 
              &\stackrel{(b)}{=} H(A) + \sum_{i=1}^{|\mathcal{U}|}P_U(u_i)\sum_{a=0}^{L-1}\!P_A(a)H(Y_{1:M_D}|A=a,U=u_i)\nonumber\\
              &\stackrel{(c)}{=}\! H(A) \!+\! \sum_{i=1}^{|\mathcal{U}|}\!P_U(u_i)\sum_{a=0}^{L-1}\!P_A(a)H_b(p_a\!*\!H_b^{-1}\big(H(X|U\!=\!u_i)\!\big)\!)\nonumber\\
              &=\sum_{i=1}^{|\mathcal{U}|}P_U(u_i)\sum_{a=0}^{L-1}\!P_A(a)\big(H_b(p_a*\tilde{p}_i)-\log P_A(a)\big) \label{eq:outputentsymm}
\end{align}
where $\displaystyle (a)$ follows because the output symbols determine $A$, $\displaystyle (b)$ follows since $A$ is independent of $X$ so that $U$ and $A$ are independent, and $(c)$ follows because $H_b(p_a*p)$ is symmetric with respect to $p=\frac{1}{2}$. Using (\ref{eq:definegw}) and (\ref{eq:outputentsymm}), we have 
\begin{align}
	g(\tilde{p})\! =\! \sum_{a=0}^{L-1}\!P_A(a)\big(H_b(p_a*\tilde{p})\!-\!\log P_A(a)\big).\label{eq:defgmix}
\end{align}
Examples of channels that are mixtures of BSCs are the dependent BSCs in Section~\ref{subsec:newsubsecforexample}, the binary erasure channel (BEC) \cite[p.~107]{GerhardITManu2017}, and additive white Gaussian noise (AWGN) channels with binary phase shift keying (BPSK) signals and symmetric (e.g., uniform) quantizers \cite[p.~108]{GerhardITManu2017}.

The convexity property (\ref{eq:convexityboundxtildeu}) carries over to channels $P_{Y_{1:M_D}|X}$ that can be decomposed into a mixture of BSCs \cite{chayat}, i.e., the function $g(\cdot)$ in (\ref{eq:defgmix}) has the property that $g(H_b^{-1}(\nu))$ is convex in $\nu$ for $0\leq\nu\leq 1$. To see this, note that $P_A(\cdot)$ is fixed and the following term in (\ref{eq:outputentsymm})$(c)$ $$\sum_{a=0}^{L\!-1}\!P_A(a)H_b(p_a*H_b^{-1}(\nu_i)))$$ where $\nu_i\!=\!H(X|U\!=\!u_i)$, is a weighted sum of convex functions of $\nu_i$ by MGL. We extend this result below in Theorem~\ref{theo:maintheorem} to show that the boundary points of $\displaystyle \mathcal{R}_1$ and $\displaystyle \mathcal{R}_2$ are achieved by channels $\displaystyle P_{\widetilde{X}|U}$ that are BSCs.

\subsection{Two Lemmas}\label{subsec:convexity}
Consider a binary-input channel $\displaystyle P_{Y_{1:M_D}|X}$. For Theorem~\ref{theo:maintheorem} below, we use the following two technical lemmas.

\begin{lemma}\label{lem:lemmaconvex}
	We have 
	\begin{align}
	H(Y_{1:M_D}|U)\geq g\Big(H_b^{-1}\big(H(X|U)\big)\Big)\text{.}\label{eq:lemma1}
	\end{align}
\end{lemma}
\begin{IEEEproof}
	Since $g(H_b^{-1}(\nu))$ is convex in $\nu$, by Jensen's inequality we have
	\begin{align*}
	&H(Y_{1:M_D}|U)= \sum_{i=1}^{\mathcal{|U|}}P_U(u_i) g\Big(H_b^{-1}\big(H_b(\tilde{p}_i)\big)\Big)\\
	&\geq g\Bigg(H_b^{-1}\Bigg(\sum_{i=1}^{\mathcal{|U|}}P_U(u_i) H_b(\tilde{p}_i)\Bigg)\Bigg) \text{.} 
	\end{align*}
\end{IEEEproof}

\begin{lemma}\label{lem:lemmaunique}
	There is a unique $\tilde{p}^*$ in the interval $\displaystyle [0,0.5]$ for which $H(X|U) = H_b(\tilde{p}^*)=\nu$.
\end{lemma}
\begin{IEEEproof}
	The function $\displaystyle H_b(\cdot)$ is strictly increasing from $0$ to $1$ in the interval $[0,0.5)$. We further have $\displaystyle 0\! \le\! H(X|U)\! \le \!H(X)\! \le \!1$.
\end{IEEEproof}

\subsection{Simplified Rate Region Characterizations}\label{subsec:simplifiedcharacterization}	
We now simplify the key-leakage-storage regions for the measurement channels $P_{\widetilde{X}|X}$ and $\displaystyle P_{Y_{1:M_D}|X}$ considered above so that a single parameter characterizes the regions.

\begin{theorem}\label{theo:maintheorem}
Suppose $P_{X|\widetilde{X}}$ is a BSC with crossover probability $p$, where $0\leq p\leq 0.5$, and $\displaystyle P_{Y_{1:M_D}|X}$ is a mixture of BSCs. The boundary points of $\displaystyle \mathcal{R}_1$ and $\displaystyle \mathcal{R}_2$ are achieved by channels $\displaystyle P_{\widetilde{X}|U}$ that are BSCs. 
\end{theorem}

\begin{IEEEproof}
Consider the boundary points of $\displaystyle \mathcal{R}_1$  
 \begin{align*}
  (R_s,R_l,R_m)\!=\! \Big(&I(U;Y_{1:M_D}\!),\nonumber\\
  &I(U;X)\!-\!I(U;Y_{1:M_D}\!),\nonumber\\
  &I(U;\widetilde{X})\!-\!I(U;Y_{1:M_D}\!)\Big)\text{.}
 \end{align*}
For a fixed $H(X|U)$, we obtain
\begin{align}
 &I(U;Y_{1:M_D})\!\leq H(Y_{1:M_D})\!-\!g\big(H_b^{-1}(H(X|U))\big)\label{eq:taga}
\end{align}
and
\begin{align}
 &I(U;X)\!-\!I(U;Y_{1:M_D})\nonumber\\
 &\!\geq\!H(X)\!-\!H(X|U)\!-\!H(Y_{1:M_D}\!)\!+\!g\big(H_b^{-1}\!(H(X|U)\!)\big)\label{eq:tagb}
\end{align}
and
\begin{align}
 &I(U;\widetilde{X})\!-\!I(U;Y_{1:M_D})\nonumber\\
 &\geq H(\widetilde{X})\! -\! H_b\Bigg(\frac{H_b^{-1}(H(X|U))-p}{1-2p}\Bigg)-H(Y_{1:M_D})\nonumber\\
 & \quad+ g\big(H_b^{-1}\!(H(X|U)\!)\big)\label{eq:tagc}
\end{align}
where we used Lemma~\ref{lem:lemmaconvex} to bound $H(Y_{1:M_D}|U)$, and the MGL result in (\ref{eq:convexityboundxtildeu}) with $\nu=H(\widetilde{X}|U)$ to bound $H(\widetilde{X}|U)$. By choosing $\displaystyle P_{U|\widetilde{X}}$ such that $\displaystyle P_{\widetilde{X}|U}$ is a BSC with crossover probability 
\begin{align}
 \tilde{x} = \frac{H_b^{-1}(H(X|U))-p}{1-2p}\label{eq:xtilderange}
\end{align}
where $\tilde{x}\in[0, 0.5]$, we achieve the right-hand sides of (\ref{eq:taga})-(\ref{eq:tagc}) since assigning $H(\widetilde{X}|U)=H_b(\tilde{x})$ achieves equality in (\ref{eq:convexityboundxtildeu}) and (\ref{eq:lemma1}) for the given channels. By Lemma~\ref{lem:lemmaunique}, this $\tilde{x}$ is the unique solution. The proof for $\displaystyle \mathcal{R}_2$ is similar.
\end{IEEEproof}

The convexity property for a BSC, used in MGL, is extended to any binary channel $P_{Y_1|X}$ by Witsenhausen in \cite{WitzenEIfDC}, by Wyner as a remark in \cite[Sec.~III]{WitzenEIfDC}, and also by Ahslwede and K\"orner in \cite{BinaryAhslwedeKörner}. Therefore, the channels $P_{Y_{1:M_D}|X}$ that can be decomposed into a mixture of \emph{binary} channels also satisfy the convexity property. This result follows because the function $g(\cdot)$ for such channels, obtained from (\ref{eq:definegw}), also consists of a constant part and a weighted sum of functions that are convex in $\nu_i$.

\begin{remark} 
	\normalfont In \cite[Theorem 1]{bizimCNS}, we claimed that for a mixture $P_{Y_{1:M_D}|X}$ of binary channels, we achieve the boundary points of $\displaystyle \mathcal{R}_1$ and $\displaystyle \mathcal{R}_2$ when $\widetilde{X}^N=X^N$ by using channels $P_{X|U}$ that are BSCs. It turns out that this claim is valid for mixtures $P_{Y_{1:M_D}|X}$ of BSCs, but not necessarily otherwise. The reason is that one cannot necessarily achieve equality in \cite[(25) and (26)]{bizimCNS}. This is illustrated in Appendix~\ref{app:nottight} for a binary asymmetric channel.   
\end{remark}

\section{Model Comparisons}\label{sec:examples}
\subsection{Hidden Source Model}
We study the GS model with a hidden binary symmetric source (BSS) such that $\displaystyle Q_X(0)\!=\!Q_X(1)\!=\!0.5$. Suppose $\displaystyle P_{\widetilde{X}|X}$ is a BSC with crossover probability $\displaystyle p_{\text{E}}$ and $\displaystyle P_{Y_{1:M_D}|X}$ consists of $M_D$ independent BSCs each with crossover probability $\displaystyle p_{\text{D}}$. The inverse channel $\displaystyle P_{X|\widetilde{X}}$ is also a BSC with crossover probability $\displaystyle p_{\text{E}}$ due to source symmetry. Due to the independence assumption for $M_D$ BSCs, the probabilities of sequences $y_{1:M_D}$ with the same Hamming weight are equal. Therefore, the decoder-output entropy is
\begin{align}
  H(Y_{1:M_D})& = \sum_{k=0}^{M_D}\Wast(\Pr\left[\sum_{m=1}^{M_D}Y_{m}\!=\!k\right]\nonumber\\
                              &\times\log_2\Bigg({M_D\choose k}\Bigg/\Pr\left[\sum_{m=1}^{M_D}Y_{m}\!=\!k\right]\Bigg)\Wast) \label{eq:decoderoutputentropy}
\end{align}
where 
\begin{align}
 \Pr\!\left[\sum_{m=1}^{M_D}\!Y_{m}\!=\!k\right]\!=\!{M_D\choose k}\Bigg(\frac{{\bar{p}}_D^{M_D\!-\!k}p_{\text{D}}^k\!+\!{\bar{p}}_D^{k}p_{\text{D}}^{M_D\!-\!k}}{2}\!\Bigg).\label{eq:decoderoutputprobability}
\end{align}
By Theorem~\ref{theo:maintheorem}, the crossover probability $\tilde{x}$, as in (\ref{eq:xtilderange}), of the BSC $P_{\widetilde{X}|U}$ is the only parameter required to characterize $\mathcal{R}_1$ for the considered source and channels. Thus, using Theorem~\ref{theo:maintheorem}, the conditional entropy $H(Y_{1:M_D}|U)$ can be calculated similarly as in (\ref{eq:decoderoutputentropy}) by using the weighted sum in (\ref{eq:decoderoutputprobability}) with the weights $\tilde{p}=\tilde{p}_1=\tilde{p}_2$ and $1-\tilde{p}$, defined in Section~\ref{sec:multimeasurecap}, instead of $\frac{1}{2}$.

\begin{figure}[t]
	\centering
	\newlength\figureheight
	\newlength\figurewidth
	\setlength\figureheight{6.47cm}
	\setlength\figurewidth{7.8cm}
%
%
\begin{tikzpicture}

\begin{axis}[%
width=0.951\figurewidth,
height=\figureheight,
at={(0\figurewidth,0\figureheight)},
scale only axis,
xmin=0,
xmax=0.7,
xlabel={Storage Rate (bits/source-bit)},
xmajorgrids,
ymin=0,
ymax=0.6,
ylabel={Privacy-leakage Rate (bits/source-bit)},
ymajorgrids,
axis background/.style={fill=white},
legend style={at={(0.00694,0.568)},anchor=south west,legend cell align=left,align=left,draw=white!15!black}
]
\addplot [color=blue,solid,line width=1.2pt,mark size=2.9pt,mark=o,mark options={solid,draw=blue}]
  table[row sep=crcr]{%
0.540750477963487	0.34635862013191\\
0.406772469046605	0.280906568160689\\
0.328242827767549	0.231349807706993\\
0.268502873383	0.191335790053045\\
0.220148693001381	0.157984511087388\\
0.179837865828707	0.129690919578486\\
0.145702021536036	0.105453860658692\\
0.116571033172733	0.0846024609190446\\
0.0916581904877612	0.0666646103769181\\
0.0704088904336984	0.0512964760889687\\
0.0524194148157495	0.0382416729044199\\
0.0373897987145246	0.0273062088358276\\
0.0250948866708886	0.0183424325983051\\
0.0153658203246374	0.0112384249241563\\
0.00807788351270999	0.00591085382320278\\
0.00314244348746562	0.00230014402165968\\
0.000501685691218501	0.00036727438788059\\
};
\addlegendentry{$\text{HSM p}_{\text{E}}\text{=0.03\&M}_\text{D}\text{=1}$};

\addplot [color=blue,solid,line width=1.2pt,mark size=2.7pt,mark=*,mark options={solid,fill=blue,draw=blue}]
  table[row sep=crcr]{%
0.292593141521995	0.0982012836904187\\
0.205245518684218	0.0793796177983022\\
0.162163450370014	0.0652704303094579\\
0.131095621979983	0.0539285386500286\\
0.106663066018212	0.0444988841042188\\
0.0866589939698033	0.0365120477195821\\
0.0699260097878643	0.0296778489105201\\
0.0557715658096228	0.0238029935559342\\
0.0437456073590217	0.0187520272481786\\
0.0335390192873594	0.0144266049426297\\
0.0249313072943151	0.0107535653829855\\
0.0177612410387647	0.00767765116006774\\
0.0119093173919989	0.00515686331941534\\
0.00728679648287089	0.00315940108238977\\
0.00382863564661162	0.00166160595710441\\
0.00148887391537855	0.000646574449572612\\
0.000237650965230429	0.000103239661892518\\
};
\addlegendentry{$\text{HSM p}_{\text{E}}\text{=0.03\&M}_\text{D}\text{=3}$};

\addplot [color=red,solid,line width=1.2pt,mark size=3.4pt,mark=diamond,mark options={solid,draw=red}]
  table[row sep=crcr]{%
0.68007704572828	0.211081452138999\\
0.525933348774048	0.179574728642137\\
0.429790250679587	0.152422579717475\\
0.354556514642145	0.128766785662363\\
0.29251816872949	0.108023438509207\\
0.240102807626671	0.0897756093811061\\
0.195266983985033	0.073714928127701\\
0.15670464190454	0.0596072818403287\\
0.123521990536187	0.0472714859103489\\
0.0950792282478991	0.0365654163999062\\
0.0709044773952649	0.0273766716529275\\
0.0506432131874496	0.0196161098982654\\
0.0340268388463691	0.0132132859362827\\
0.0208523676422351	0.00811318744850348\\
0.0109689533999643	0.00427389212719032\\
0.00426888622261268	0.00166490139019515\\
0.000681670044192118	0.000265992016609762\\
};
\addlegendentry{$\text{HSM p}_{\text{E}}\text{=0.10\&M}_\text{D}\text{=1}$};

\addplot [color=red,solid,line width=1.2pt,mark size=2.7pt,mark=diamond*,mark options={solid,fill=red,draw=red}]
  table[row sep=crcr]{%
0.528516521533553	0.0595209279442715\\
0.396959748923541	0.0506011287916308\\
0.320295670957144	0.0429279999950321\\
0.262041077114314	0.0362513481345321\\
0.214896838920496	0.0304021087002129\\
0.175587351973686	0.0252601537281213\\
0.142289013697967	0.0207369578406345\\
0.113862861588776	0.0167655015245643\\
0.0895445358645803	0.0132940312387427\\
0.0687958337440391	0.0102820218960463\\
0.0512252742722244	0.00769746852988701\\
0.0365421212375557	0.0055150179483715\\
0.0245282057311652	0.00371465282107886\\
0.0150199327029116	0.00228075250917992\\
0.00789648129074361	0.00120142001796963\\
0.00307198879224635	0.000468003959828822\\
0.000490447469055133	7.47694414727773e-05\\
};
\addlegendentry{$\text{HSM p}_{\text{E}}\text{=0.10\&M}_\text{D}\text{=3}$};

\addplot [color=blue,dashed,line width=1.2pt,mark size=2.9pt,mark=square,mark options={solid,draw=blue}]
  table[row sep=crcr]{%
0.540750477963487	0.540750477963487\\
0.406772469046605	0.406772469046605\\
0.328242827767549	0.328242827767549\\
0.268502873383	0.268502873383\\
0.220148693001381	0.220148693001381\\
0.179837865828707	0.179837865828707\\
0.145702021536036	0.145702021536036\\
0.116571033172733	0.116571033172733\\
0.0916581904877612	0.0916581904877612\\
0.0704088904336984	0.0704088904336984\\
0.0524194148157495	0.0524194148157495\\
0.0373897987145246	0.0373897987145246\\
0.0250948866708886	0.0250948866708886\\
0.0153658203246373	0.0153658203246373\\
0.00807788351270994	0.00807788351270994\\
0.00314244348746562	0.00314244348746562\\
0.000501685691218501	0.000501685691218501\\
};
\addlegendentry{$\text{VSM p}_{\text{E}}\text{=0.03\&M}_\text{D}\text{=1}$};

\addplot [color=blue,dashed,line width=1.2pt,mark size=3.4pt,mark=square*,mark options={solid,fill=blue,draw=blue}]
  table[row sep=crcr]{%
0.195103431989355	0.195103431989355\\
0.141428811640303	0.141428811640303\\
0.113572769265464	0.113572769265464\\
0.0926844078316878	0.0926844078316878\\
0.0758843315068088	0.0758843315068088\\
0.0619291586471372	0.0619291586471372\\
0.0501388712442293	0.0501388712442293\\
0.040093139464171	0.040093139464171\\
0.0315118168395331	0.0315118168395331\\
0.0241986214477838	0.0241986214477838\\
0.0180113150681461	0.0180113150681461\\
0.0128445717558568	0.0128445717558568\\
0.00861953716117769	0.00861953716117769\\
0.00527718988933201	0.00527718988933201\\
0.0027740025663483	0.0027740025663483\\
0.00107907521112394	0.00107907521112394\\
0.000172267250183045	0.000172267250183045\\
};
\addlegendentry{$\text{VSM p}_{\text{E}}\text{=0.03\&M}_\text{D}\text{=3}$};

\end{axis}
\end{tikzpicture}%
	\caption{Storage-leakage projection of the boundary triples for the GS model with $\displaystyle p_{\text{D}}\!=\!0.10$.} \label{fig:xyaxisproject}
\end{figure}
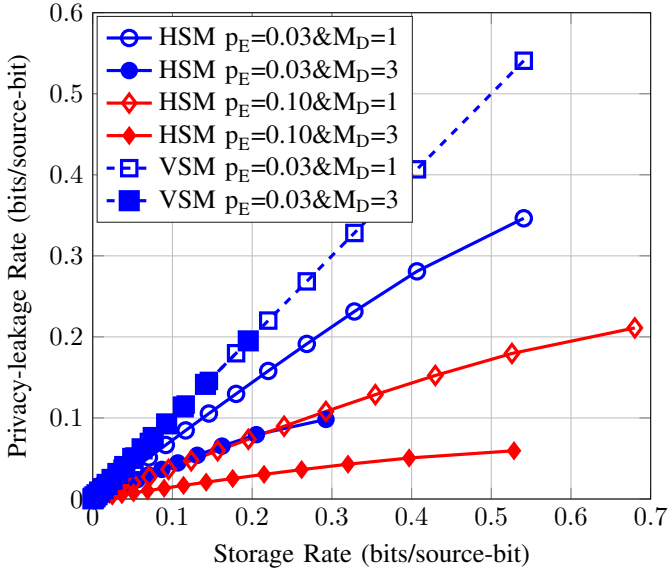

\subsection{Mismatched Code Design}
The encoder, e.g., a hardware manufacturer (for PUFs) or a trusted entity (for biometrics), models the source as visible or hidden, and a code is then constructed for the assumed model. Therefore, the assumed model determines the performance of the actual system. In the literature, the physical and biometric identifiers are modeled by the VSM; see, e.g., \cite{IgnaTrans,LaiTrans,FuzzyCommitment,Dodis2008fuzzy}. We first illustrate that treating the HSM as if it were a VSM might give pessimistic privacy-leakage rate results for $M_D\geq 1$ and over-optimistic secret-key and storage rate results for $M_D\!>\!1$, which results in unnoticed secrecy leakage and reduced reliability.

Consider the crossover probabilities $\displaystyle p_{\text{E}}\!\in\!\{\!0.03, 0.10\!\}$, which are realistic values for biometric \cite{IgnaTrans} and physical identifiers \cite{bizimpaper}. We fix the crossover probability of $\displaystyle P_{Y_{i}|X}$ for $\displaystyle i\!=\!1,2,\ldots,M_D$ to $\displaystyle p_{\text{D}}\!=\!0.10$. For the supposed VSM, $\widetilde{X}^N$ is mistakenly considered to be a noise-free source, i.e., $p_{\text{E}}^{VSM}\!=\!0$, and the corresponding decoder-output channel $P_{Y_{1:M_D}|\widetilde{X}}^{VSM}$ consists of $M_D$ independent BSCs each with crossover probability $\displaystyle p_{\text{E}}*p_{\text{D}}$ because $P_{Y|\widetilde{X}}$ is estimated from identifier measurements. However, the HSM considers an encoder measurement through a BSC with crossover probability $p_{\text{E}}$ and $M_D$ independent decoder measurements through BSCs, each with crossover probability $p_D$. Therefore, the HSM results in a conditional probability distribution $P_{Y_{1:M_D}|\widetilde{X}}$ that is different from the supposed VSM distribution $P_{Y_{1:M_D}|\widetilde{X}}^{VSM}$ for $M_D\!>\!1$ and in a key-leakage-storage region $\mathcal{R}_1$ that is different from the supposed VSM region $\mathcal{R}_1^{VSM}$ for $M_D\!\geq\!1$. We next illustrate the rate regions for different numbers of encoder and decoder measurements with $p_{\text{E}}$ and $p_{\text{D}}$ values given above.

\subsection{Single Encoder and Multiple Decoder Measurements}
The projections of the boundary triples $\displaystyle (R_s,R_l,R_m)$ for the HSM and VSM onto the $\displaystyle (R_m,R_l)$-plane and onto the $\displaystyle (R_m,R_s)$-plane are depicted in Fig.~\ref{fig:xyaxisproject} and Fig.~\ref{fig:xzaxisproject}, respectively, for different crossover probabilities at the encoder and different numbers of measurements at the decoder. Every marker on each curve corresponds to the evaluation of the rate-region boundaries for a fixed crossover probability $\tilde{x}$ given in (\ref{eq:xtilderange}), so Figs.~\ref{fig:xyaxisproject} and ~\ref{fig:xzaxisproject} should be considered jointly for analysis. Recall from (\ref{eq:rateregiongenerated}) that any smaller $R_s$ and greater $R_l$ and $R_m$ than the boundary triples are achievable.

\begin{figure}[t]
	\centering
	\setlength\figureheight{6.67cm}
	\setlength\figurewidth{7.8cm}
%
%
\begin{tikzpicture}

\begin{axis}[%
width=0.951\figurewidth,
height=\figureheight,
at={(0\figurewidth,0\figureheight)},
scale only axis,
xmin=0,
xmax=0.7,
xlabel={Storage Rate (bits/source-bit)},
xmajorgrids,
ymin=0,
ymax=0.9,
ylabel={Secret-key Rate (bits/source-bit)},
ymajorgrids,
axis background/.style={fill=white},
legend style={at={(0.43,0.577)},anchor=south west,legend cell align=left,align=left,draw=white!15!black}
]
\addplot [color=blue,solid,line width=1.2pt,mark size=3.5pt,mark=o,mark options={solid,draw=blue}]
  table[row sep=crcr]{%
0.540750477963487	0.459249522036513\\
0.406772469046605	0.398835673121819\\
0.328242827767549	0.344312253077974\\
0.268502873383	0.295027309552897\\
0.220148693001381	0.250490441711255\\
0.179837865828707	0.210321829454892\\
0.145702021536036	0.174220932735684\\
0.116571033172733	0.141946226895993\\
0.0916581904877612	0.113301530127717\\
0.0704088904336984	0.088126473358126\\
0.0524194148157495	0.0662896859535578\\
0.0373897987145246	0.047683828505748\\
0.0250948866708886	0.0322219240736192\\
0.0153658203246374	0.0198346311702754\\
0.00807788351270999	0.0104682214536365\\
0.00314244348746562	0.0040831025247261\\
0.000501685691218501	0.000652778313579738\\
};
\addlegendentry{$\text{HSM p}_{\text{E}}\text{=0.03\&M}_\text{D}\text{=1}$};

\addplot [color=blue,solid,line width=1.2pt,mark size=2.5pt,mark=*,mark options={solid,fill=blue,draw=blue}]
  table[row sep=crcr]{%
0.292593141521995	0.707406858478005\\
0.205245518684218	0.600362623484206\\
0.162163450370014	0.51039163047551\\
0.131095621979983	0.432434560955914\\
0.106663066018212	0.363976068694424\\
0.0866589939698033	0.303500701313796\\
0.0699260097878643	0.249996944483856\\
0.0557715658096228	0.202745694259104\\
0.0437456073590217	0.161214113256456\\
0.0335390192873594	0.124996344504465\\
0.0249313072943151	0.0937777934749922\\
0.0177612410387647	0.0673123861815079\\
0.0119093173919989	0.0454074933525089\\
0.00728679648287089	0.0279136550120419\\
0.00382863564661162	0.0147174693197348\\
0.00148887391537855	0.00573667209681317\\
0.000237650965230429	0.00091681303956781\\
};
\addlegendentry{$\text{HSM p}_{\text{E}}\text{=0.03\&M}_\text{D}\text{=3}$};

\addplot [color=red,solid,line width=1.2pt,mark size=4.5pt,mark=diamond,mark options={solid,draw=red}]
  table[row sep=crcr]{%
0.68007704572828	0.31992295427172\\
0.525933348774048	0.279674793394376\\
0.429790250679587	0.242764830165936\\
0.354556514642145	0.208973668293752\\
0.29251816872949	0.178120965983146\\
0.240102807626671	0.150056887656928\\
0.195266983985033	0.124655970286687\\
0.15670464190454	0.101812618164186\\
0.123521990536187	0.0814377300792912\\
0.0950792282478991	0.0634561355439253\\
0.0709044773952649	0.0478046233740424\\
0.0506432131874496	0.0344304140328229\\
0.0340268388463691	0.0232899718981387\\
0.0208523676422351	0.0143480838526777\\
0.0109689533999643	0.00757715156638217\\
0.00426888622261268	0.00295665978957904\\
0.000681670044192118	0.000472793960606122\\
};
\addlegendentry{$\text{HSM p}_{\text{E}}\text{=0.10\&M}_\text{D}\text{=1}$};

\addplot [color=red,solid,line width=1.2pt,mark size=4.5pt,mark=diamond*,mark options={solid,fill=red,draw=red}]
  table[row sep=crcr]{%
0.528516521533553	0.471483478466447\\
0.396959748923541	0.408648393244883\\
0.320295670957144	0.35225940988838\\
0.262041077114314	0.301489105821583\\
0.214896838920496	0.25574229579214\\
0.175587351973686	0.214572343309913\\
0.142289013697967	0.177633940573754\\
0.113862861588776	0.144654398479951\\
0.0895445358645803	0.115415184750897\\
0.0687958337440391	0.0897395300477852\\
0.0512252742722244	0.0674838264970829\\
0.0365421212375557	0.0485315059827168\\
0.0245282057311652	0.0327886050133426\\
0.0150199327029116	0.0201805187920012\\
0.00789648129074361	0.0106496236756028\\
0.00307198879224635	0.00415355721994537\\
0.000490447469055133	0.000664016535743106\\
};
\addlegendentry{$\text{HSM p}_{\text{E}}\text{=0.10\&M}_\text{D}\text{=3}$};

\addplot [color=blue,dashed,line width=1.2pt,mark size=4.5pt,mark=square,mark options={solid,draw=blue}]
  table[row sep=crcr]{%
0.540750477963487	0.459249522036513\\
0.406772469046605	0.398835673121819\\
0.328242827767549	0.344312253077974\\
0.268502873383	0.295027309552897\\
0.220148693001381	0.250490441711255\\
0.179837865828707	0.210321829454892\\
0.145702021536036	0.174220932735684\\
0.116571033172733	0.141946226895993\\
0.0916581904877612	0.113301530127717\\
0.0704088904336984	0.088126473358126\\
0.0524194148157495	0.0662896859535578\\
0.0373897987145246	0.047683828505748\\
0.0250948866708886	0.0322219240736192\\
0.0153658203246373	0.0198346311702755\\
0.00807788351270994	0.0104682214536365\\
0.00314244348746562	0.0040831025247261\\
0.000501685691218501	0.000652778313579738\\
};
\addlegendentry{$\text{VSM p}_{\text{E}}\text{=0.03\&M}_\text{D}\text{=1}$};

\addplot [color=blue,dashed,line width=1.2pt,mark size=2.5pt,mark=square*,mark options={solid,fill=blue,draw=blue}]
  table[row sep=crcr]{%
0.195103431989355	0.804896568010645\\
0.141428811640303	0.664179330528121\\
0.113572769265464	0.55898231158006\\
0.0926844078316878	0.470845775104209\\
0.0758843315068088	0.394754803205827\\
0.0619291586471372	0.328230536636462\\
0.0501388712442293	0.269784083027491\\
0.040093139464171	0.218424120604555\\
0.0315118168395331	0.173447903775945\\
0.0241986214477838	0.13433674234404\\
0.0180113150681461	0.100697785701161\\
0.0128445717558568	0.0722290554644158\\
0.00861953716117769	0.0486972735833301\\
0.00527718988933201	0.0299232616055808\\
0.0027740025663483	0.0157721023999982\\
0.00107907521112394	0.00614647080106778\\
0.000172267250183045	0.000982196754615194\\
};
\addlegendentry{$\text{VSM p}_{\text{E}}\text{=0.03\&M}_\text{D}\text{=3}$};

\end{axis}
\end{tikzpicture}%
	\caption{Storage-key projection of the boundary triples for the GS model with $\displaystyle p_{\text{D}}\!=\!0.10$.} \label{fig:xzaxisproject}
\end{figure}
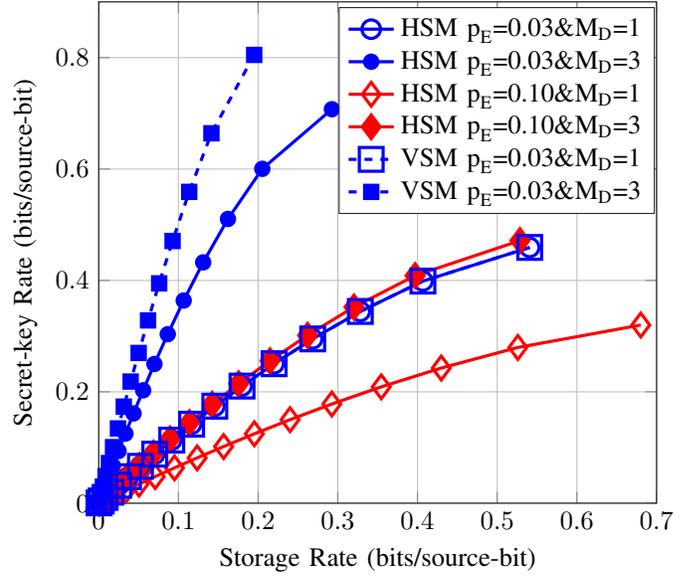

At the highest storage-leakage points $\displaystyle (R_m^*,R_l^*)$ in Fig.~\ref{fig:xyaxisproject}, one achieves the maximum secret-key rates $\displaystyle R_s^*$, which corresponds to the highest points in Fig.~\ref{fig:xzaxisproject}. Moreover, Fig.~\ref{fig:xyaxisproject} shows that if $M_D\!=\!1$, for the supposed VSM the privacy-leakage and storage rates are equal, and are also equal to the storage rate for the HSM, and the supposed VSM gives pessimistic privacy-leakage rate results. Fig.~\ref{fig:xzaxisproject} shows that increasing the number of decoder measurements increases the maximum secret-key rate $\displaystyle R_s^*$ for the HSM and supposed VSM. The $R_s^*$ of the HSM and supposed VSM are equal if $M_D\!=\!1$, but the supposed VSM gives over-optimistic secret-key and storage rate results for $M_D\!>1$. These comparisons show that designing a code for the supposed VSM can lead to substantial secrecy leakage, which violates (\ref{eq:secrecyconst}),  and reliability reduction, which violates (\ref{eq:reliabilityconst}).

Consider, for instance, the parameters $\big(p_{\text{E}}\!=\!0.03$, $p_{\text{D}}\!=\!0.10$, $M_D\,{=}\,35\big)$. For the GS and CS models with the HSM, $R_l^*$ is approximately $3\times10^{-9}$ bits/source-bit. The privacy-leakage rate can thus be made small for both models with multiple decoder measurements. $R_m^*$ is approximately $0.194$ bits/source-bit for the GS model, which is smaller than the $0.541$ bits/source-bit obtained for $M_D\!=\!1$. We remark that less storage decreases the hardware cost. It is not possible for the CS model to give such a small storage rate with multiple decoder measurements since the key is independent of the hidden source. Independence of the key results in an additional storage rate, equal to $R_s^*$ that is non-decreasing with respect to $M_D$, to reliably reconstruct the secret key at the decoder.
 
One may build intuition about the gains achieved by having multiple decoder measurements as follows. Since the decoder sees $\displaystyle M_D$ noisy versions $\displaystyle Y_{1:M_D}$ of the same hidden source symbol $\displaystyle X$, it can ``combine'' the measurements to form a less noisy equivalent channel. This is entirely similar to using maximal ratio combining to obtain a sufficient statistic about a symbol that is transmitted several times over an AWGN channel. The resulting gain may thus be interpreted as a diversity gain.
 
Figs.~\ref{fig:xyaxisproject} and ~\ref{fig:xzaxisproject} further show that increasing $\displaystyle p_{\text{E}}$ decreases $\displaystyle R_s^*$ and $\displaystyle R_l^*$, and increases $\displaystyle R_m^*$ for the HSM. For instance, consider the HSM, and fix $M_D=1$ and the secret-key rate to its maximum $R_s^*=0.320$ bits/source-bit achieved for $p_{\text{E}}=0.10$. The storage rate for $p_{\text{E}}=0.03$ is approximately $56.2\%$ less than the storage rate for $p_{\text{E}}=0.10$ and their privacy-leakage rates are equal. Therefore, more reliable encoder-output channels $P_{\widetilde{X}|X}$, i.e., channels with smaller $p_{\text{E}}$ values, achieve better storage rates. Similarly, we can show that more reliable decoder-output channels $P_{Y_{1:M_D}|X}$, i.e., channels with smaller $p_{\text{D}}$ values, improve the rate triples (see also \cite[Fig. 5]{bizimCNS}) due to the independence assumption for encoder and decoder measurements. 

\begin{remark} 
	\normalfont One can alternatively consider encoder and decoder measurements through a broadcast channel. An unreliable channel at the encoder might be desirable for this case if the decoder-output channel is unreliable, since such correlations might allow less storage and privacy-leakage, and greater secret-key rates.
\end{remark}

\subsection{Multiple Encoder and Decoder Measurements}
Consider the general case with $M_E\!\geq\!1$ measurements $\widetilde{X}^N_{1:M_E}\!=\!{[\widetilde{X}_1\widetilde{X}_2\ldots\widetilde{X}_{M_E}]}^N$ of the hidden source $X^N$ at the encoder for the GS model with the HSM. Suppose each encoder-output channel $P_{\widetilde{X}_i|X}$ for $i\!=\!1,2,\ldots,M_E$ is an independent BSC with crossover probability $p_{\text{E}}$. The maximum secret-key rate $R_s^*$ is achieved by choosing $U\!=\!\widetilde{X}_{1:M_E}$ for $\mathcal{R}_1$. We list the $\displaystyle (R_s^*,R_l^*,R_m^*)$ points in Table~\ref{tab:multisecretleakagestorage} for different numbers of encoder and decoder measurements with $p_{\text{E}}\!=\! 0.03$ and $p_{\text{D}}\!=\!0.10$. Table~\ref{tab:multisecretleakagestorage} shows that the storage rates for multiple encoder measurements can be greater than $1$ bit/source-bit, which cannot be the case for a single encoder measurement. Increasing the number $M_E$ of encoder measurements to increase the secret-key rate, as listed in Table~\ref{tab:multisecretleakagestorage}, can therefore come at a large cost of storage and can increase the privacy-leakage rate.

\begin{table}[t]
	\centering
	\caption{Key-leakage-storage $(R_s^*,R_l^*,R_m^*)$ rate points for the GS model with the HSM for $\displaystyle p_{\text{E}}\!=\!0.03$ and $\displaystyle p_{\text{D}}\!=\!0.10$.}
	\begin{tabular}{ |c|c| }
		\hline
		{\cellcolor{blue!25}$\left(M_E,M_D\right)$} & \cellcolor{blue!25} $(R_s^*,R_l^*,R_m^*)$ bits/source-bit \\
		\hline
		$(1,1)$ &  $(0.459,0.346,0.541)$\\
		\hline
		$(1,3)$ &  $(0.707,0.098,0.293)$\\
		\hline
		$(3,1)$ & $(0.525,0.458,1.041)$\\
		\hline
		$(3,3)$ &  $(0.849,0.134,0.717)$\\
		\hline
	\end{tabular}\label{tab:multisecretleakagestorage}
\end{table}

\section{Achievability Proofs}\label{sec:achProofs}
\subsection{Achievability Proof for Theorem~\ref{theo:regiongenerated}}
\subsubsection{Overview}
We choose the conditional probabilities $\displaystyle P_{U|\widetilde{X}}(u|\tilde{x})$ for all $\displaystyle u\!\in\!\mathcal{U}$ and $\displaystyle \tilde{x}\!\in\!\mathcal{\widetilde{X}}$. We randomly and independently generate about $\displaystyle 2^{NI(U;\widetilde{X})}$ sequences $\displaystyle u^N(m,s)$ for $\displaystyle m\!=\!1,\ldots,2^{NR_m}$ and $\displaystyle s\!=\!1,\ldots,2^{NR_s}$. Consider approximately $\displaystyle 2^{N(I(U;\widetilde{X})-I(U;Y))}$ storage labels $m$ and $2^{NI(U;Y)}$ key labels $s$, which can be considered as bins. The encoder finds a $\displaystyle u^N(m,s)$ sequence that is jointly typical with the observed measurement $\displaystyle \tilde{x}^N$ of the source $x^N$. It then publicly sends the storage label $m$ to the decoder. The decoder sees another measurement $y^N$ of the source and it determines the unique $\displaystyle u^N(m,\hat{s})$ that is jointly typical with $y^N$. Using standard arguments, one can show that the error probability $\displaystyle \Pr[S\!\ne\!\hat{S}]\!\rightarrow\!0$ as $\displaystyle N\!\rightarrow\!\infty$. The secrecy-leakage rate is negligible if there is no error. The privacy-leakage rate is approximately $\displaystyle I(U;X)\!-\!I(U;Y)$, which requires a different analysis than in \cite{IgnaTrans}. 

\subsubsection{Proof}
Fix $\displaystyle P_{U|\widetilde{X}}$. Randomly and independently generate codewords $\displaystyle u^N(m,s)$, $m\!=\!1,\ldots,2^{NR_m}$, $s\!=\!1,\ldots,2^{NR_s}$ according to $\prod_{i=1}^{N}P_U(u_i)$, where
\begin{align}
P_U(u_i) = \sum_{(\tilde{x},x)\in\mathcal{\widetilde{X}}\times\mathcal{X}}P_{U|\widetilde{X}}(u_i|\tilde{x})P_{\widetilde{X}|X}(\tilde{x}|x)Q_X(x).
\end{align}
These codewords define the codebook 
\begin{align*}
\mathcal{C} = \{u^N(m,s), m=1,\ldots,2^{NR_m}, s=1,\ldots,2^{NR_s}\}
\end{align*}
and we denote the random codebook by
\begin{align}
\mathcal{\tilde{C}}={\{U^N(m,s)\}}^{(2^{NR_m},2^{NR_s})}_{(m,s)=(1,1)}.
\end{align}
Let $\displaystyle 0\!<\!\epsilon^{\prime}\!<\!\epsilon$.  

\emph{Encoding}: Given $\displaystyle \tilde{x}^N$, the encoder looks for a codeword that is jointly typical with $\tilde{x}^N$, i.e., $\displaystyle (u^N(m,s),\tilde{x}^N)\!\in\!\mathcal{T}_{\epsilon^{\prime}}^{N}(P_{U\widetilde{X}})$. If there is one or more such codeword, then the encoder chooses one of them and puts out $(m,s)$.  If there is no such codeword, set $m\!=\!s\!=\!1$. The encoder publicly stores the label $m$.

\emph{Decoding}: The decoder puts out $\hat{s}$ if there is a unique key label $\hat{s}$ that satisfies the typicality check $\displaystyle (u^N(m,\hat{s}),y^N)\!\in\!\mathcal{T}_{\epsilon}^{N}(P_{UY})$; otherwise, it sets $\displaystyle \hat{s}\!=\!1$. 

\emph{Error Probability}: Define the error events

\begin{flalign*}
&E_{1} = \big\{(U^N(m,s),\widetilde{X}^N)\not\in \mathcal{T}_{\epsilon^{\prime}}^{N}(P_{U\widetilde{X}}) \text{ for all }\nonumber\\
&\qquad\quad (m,s)\!\in\![1\!:\!2^{NR_m}]\times[1\!:\!2^{NR_s}]\big\}\nonumber\\
&E_{2} = \big\{(U^N(M,s),\widetilde{X}^N,Y^N)\not\in \mathcal{T}_{\epsilon}^{N}(P_{U\widetilde{X}Y}) \text{ for all }\nonumber\\ &\qquad\quad s\!\in\![1\!:\!2^{NR_s}]\big\}\nonumber\\
&E_{3} = \big\{(U^N(M,s^{\prime}),Y^N)\in \mathcal{T}_{\epsilon}^{N}(P_{UY}) \text{ for some } s^{\prime}\!\ne\!S\big\}.&&
\end{flalign*}
and the overall error event $E=\cup_{i=1}^3 E_i$. Using the union bound, we have
\begin{align}
\Pr[E]\! \leq\! \Pr[E_1]\!+\!\Pr[E_1^{c}\cap E_2]\! +\! \Pr[E_3]. 
\end{align}

$\displaystyle \Pr[E_1]$ is small with large $\displaystyle N$ and small $\epsilon^{\prime}$ if
\begin{align}
R_m\!+\!R_s\!>\!I(U;\widetilde{X})\!+\!\delta(\epsilon^{\prime})  
\end{align}
where $\delta(\epsilon^{\prime})$ is small with small $\displaystyle \epsilon^{\prime}$ (see the covering lemma \cite[Lemma 3.3]{Elgamalbook}). 

Note that the event $\{\widetilde{X}^N\!=\!\tilde{x}^N\!,\,\!U^N\!=\!u^N\!\}$ implies $Y^N\!\sim\!\prod_{i=1}^{N}P_{Y|\widetilde{X}}(y_i|\tilde{x}_i)$. By the conditional typicality lemma \cite[Section 2.5]{Elgamalbook}, we obtain that $ \displaystyle \Pr[E_1^{c}\cap E_2]$ is small with large $\displaystyle N$. 

Due to symmetry in the code generation, we can set $M\!=\!1$ and have 
\begin{flalign}
\Pr[E_3]\! =\! \Pr[(U^N(1,s^{\prime}),Y^N\!)\!\in\! \mathcal{T}_{\epsilon}^{N}(P_{UY}\!)\text{ for some } s^{\prime}\!\ne\!S].\nonumber
\end{flalign}
Using the packing lemma \cite[Lemma 3.1]{Elgamalbook}, we find that $\Pr[E_3]$ is small with large $\displaystyle N$ and small $\epsilon$ if
\begin{align}
R_s\!<\!I(U;Y)\!-\!\delta(\epsilon)
\end{align}
where $\displaystyle \delta(\epsilon)$ is small with small $\displaystyle \epsilon$. 

We therefore define some $\displaystyle\delta_1$ and $\delta_2$, where $\displaystyle\delta_2\!>\!\delta(\epsilon)$ and $\displaystyle\delta_1\!>\!\delta(\epsilon^{\prime})\!+\!\delta_2$, that are small with small $\displaystyle \epsilon$ and some $\displaystyle\delta^{\prime}$ that is small with large $\displaystyle N$ and small $\displaystyle \epsilon$ such that  
\begin{flalign}
&\Pr[E]\leq \displaystyle\delta^{\prime}\label{eq:perror}\\
&R_m = I(U;\widetilde{X})-I(U;Y)+\delta_1\label{eq:r_mchoose}\\
&R_s = I(U;Y) - \delta_2\label{eq:r_schoose}.
\end{flalign}

We first establish bounds on the secrecy-leakage, secret-key, privacy-leakage, and storage rates averaged over the random codebook $\mathcal{\tilde{C}}$ and then we show that there exists a codebook satisfying (\ref{eq:reliabilityconst})-(\ref{eq:storageconst}). In the following, $U^N$ represents $U^N(M,S)$. 

\emph{Secrecy-leakage Rate}: Observe that
\begin{align}
&H(MS|\mathcal{\tilde{C}})\!\stackrel{(a)}{=} H(U^NMS|\mathcal{\tilde{C}})\nonumber\\
&\!\geq H(U^N|\mathcal{\tilde{C}})\nonumber\\
&\!=H(U^N\widetilde{X}^N|\mathcal{\tilde{C}})-H(\widetilde{X}^N|U^N\mathcal{\tilde{C}})\nonumber\\
&\!\stackrel{(b)}{\geq}NH(\widetilde{X})\!-\!H(\widetilde{X}^N|U^N\mathcal{\tilde{C}})\nonumber\\
&\!\stackrel{(c)}{\geq}NH(\widetilde{X})\!-\!N(H(\widetilde{X}|U)+\delta_{\epsilon})\nonumber
\end{align}
\begin{align}
&\!=N(I(U;\widetilde{X})\!-\!\delta_{\epsilon})\nonumber\\
&\!\stackrel{(d)}{=}N(R_m\!+\!R_s\!-\delta_1\!+\!\delta_2\!-\!\delta_{\epsilon})\label{eq:HMSlower}
\end{align}
where \\
$(a)$ follows because, given the codebook, $MS$ determines $U^N$,\\
$(b)$ follows because $\widetilde{X}^N$ is independent of the codebook,\\
$(c)$ follows by using \cite[Lemma 4]{Kittipong_a} for $\delta_{\epsilon}$ that is small with small $\epsilon$,\\
$(d)$ follows by (\ref{eq:r_mchoose}) and (\ref{eq:r_schoose}). 

Using (\ref{eq:HMSlower}), we obtain
\begin{align}
&\frac{1}{N}I(S;M|\mathcal{\tilde{C}})\!=\! \frac{1}{N}(H(S|\mathcal{\tilde{C}})\!+\!H(M|\mathcal{\tilde{C}})\!-\!H(MS|\mathcal{\tilde{C}}))\nonumber\\
&\!\leq\!\frac{1}{N}(NR_s\!+\!NR_m\!-\!H(MS|\mathcal{\tilde{C}}))\nonumber\\
&\!\leq\delta_1\!-\!\delta_2\!+\!\delta_{\epsilon}
\label{eq:secrecyleakageresult}
\end{align}
which is small with small $\displaystyle \epsilon$.

\emph{Key Uniformity}: We have
\begin{align}
&\frac{1}{N}H(S|\mathcal{\tilde{C}}) \geq \frac{1}{N}(H(MS|\mathcal{\tilde{C}})\!-\!H(M|\mathcal{\tilde{C}}))\nonumber\\
&\stackrel{(a)}{\geq}R_s\!-\delta_1\!+\!\delta_2\!-\!\delta_{\epsilon}.\label{eq:keyunifresult}
\end{align}
where $(a)$ follows by (\ref{eq:HMSlower}). 

\emph{Privacy-leakage Rate}: First, consider  
\begin{align}
H&(M|X^N\mathcal{\tilde{C}})=H(M\widetilde{X}^N|X^N\mathcal{\tilde{C}})\!-\!H(\widetilde{X}^N|MX^N\mathcal{\tilde{C}})\nonumber\\
\stackrel{(a)}{\geq}& H(\widetilde{X}^N|X^N)\!-\!H(\widetilde{X}^N|MX^N\mathcal{\tilde{C}})\nonumber\\
\stackrel{(b)}{=}& H(\widetilde{X}^N|X^N)\!-\!H(\widetilde{X}^NS|MX^N\mathcal{\tilde{C}})\nonumber\\
\stackrel{(c)}{\geq}& NH(\widetilde{X}|X)\!-\! H(S|MX^N\mathcal{\tilde{C}})\!-\!H(\widetilde{X}^N|X^NU^N\mathcal{\tilde{C}})\nonumber\\
\stackrel{(d)}{=}& NH(\widetilde{X}|X)\!-\!H(S|MX^NY^N\hat{S}\mathcal{\tilde{C}})\!-\!H(\widetilde{X}^N|X^NU^N\mathcal{\tilde{C}})\nonumber\\
\stackrel{(e)}{\geq}&NH(\widetilde{X}|X)\!-\!\Pr[E]\log\!|\mathcal{S}|\!-\!H_b(\Pr[E])\nonumber\\
&\!-\!H(\widetilde{X}^N|X^NU^N\mathcal{\tilde{C}})\nonumber\\
\stackrel{(f)}{=}&NH(\widetilde{X}|X)\!-\!N\delta^{\prime\prime}\!-\!H(\widetilde{X}^N|X^NU^N\mathcal{\tilde{C}})\nonumber\\
\stackrel{(g)}{\geq}&NH(\widetilde{X}|X)\!-\!N\delta^{\prime\prime}\!-\!N(H(\widetilde{X}|XU)\!+\!\delta_{\epsilon}^{\prime})\nonumber\\
=&N(I(U;\widetilde{X}|X)\!-\!(\delta^{\prime\prime}\!+\!\delta_{\epsilon}^{\prime}))
\label{eq:leakstep1}
\end{align}
where\\ 
$(a)$ follows because $\mathcal{\tilde{C}}$ is independent of $\widetilde{X}^NX^N$,\\
$(b)$ follows because, given the codebook, $\widetilde{X}^N$ determines $S$,\\
$(c)$ follows since, given the codebook, $MS$ determines $U^n$,\\
$(d)$ follows by the Markov chain $SMU^N\!-\!\widetilde{X}^N\!-\!X^N\!-\!Y^N$, \\
$(e)$ follows from Fano's inequality,\\
$(f)$ follows by using $|\mathcal{S}|\!\leq\!|\mathcal{\widetilde{X}}|^N$ and defining a parameter $\delta^{\prime\prime}$ that is small with large $\displaystyle N$ and small $\displaystyle \epsilon$ due to (\ref{eq:perror}),\\
$(g)$ follows by using \cite[Lemma 4]{Kittipong_a} for $\delta_{\epsilon}$ that is small with small $\epsilon$.

Using (\ref{eq:leakstep1}), we have
\begin{align}
\frac{1}{N}&I(X^N;M|\mathcal{\tilde{C}})\!=\!\frac{1}{N}(H(M|\mathcal{\tilde{C}})\!-\!H(M|X^N\mathcal{\tilde{C}}))\nonumber\\
\leq& R_m\!-\!(I(U;\widetilde{X}|X)\!-\!(\delta^{\prime\prime}\!+\!\delta_{\epsilon}^{\prime}))\nonumber\\
\stackrel{(a)}{=}&R_m\!-\!(H(U|X)\!-\!H(U|\widetilde{X})\!-\!(\delta^{\prime\prime}\!+\!\delta_{\epsilon}^{\prime}))\nonumber
\end{align}
\begin{align}            
\stackrel{(b)}{=}&I(U;X)\!-\!I(U;Y)\!+\!\delta^{\prime\prime}\!+\!\delta_{\epsilon}^{\prime}\!+\!\delta_1\label{eq:privacyleaklast}
\end{align}
where $(a)$ follows by the Markov chain $\displaystyle U\!-\!\widetilde{X}\!-\!X$ and $(b)$ follows by (\ref{eq:r_mchoose}). 

\emph{Storage Rate}: Using (\ref{eq:r_mchoose}), we have
\begin{align}
&\frac{1}{N}H(M|\mathcal{\tilde{C}})\leq R_m =I(U;\widetilde{X})-I(U;Y)+\delta_1.
\end{align}

Applying the selection lemma \cite[Lemma 2.2]{Blochbook} to these results, there exists a codebook for the GS model that approaches the key-leakage-storage triple
\begin{align*}
(R_s,R_l,R_m)=\big(&I(U;Y),\nonumber\\
&I(U;X)\!-\!I(U;Y),\nonumber\\
&I(U;\widetilde{X})\!-\!I(U;Y)\big).
\end{align*}

\subsection{Achievability Proof for Theorem~\ref{theo:regionchosen}} 
\subsubsection{Overview}
We use the achievability proof of the GS model in combination with a one-time pad to conceal the embedded secret key $S$ by the key $\displaystyle S^{\prime}$ generated by the GS model. The embedded key $\displaystyle S$ is uniformly distributed and independent of other random variables. The secret-key and privacy-leakage rates do not change, but the storage rate $\displaystyle I(U;\widetilde{X})$ is approximately the sum of the storage and secret-key rates of the GS model.

\subsubsection{Proof}
Suppose $S$ has the same cardinality as $\displaystyle S^{\prime}$, i.e., $|\mathcal{S}|\!=\!|\mathcal{S^{\prime}}|$. We use the codebook, encoder, and decoder of the GS model and add the masking layer (one-time pad) approach of \cite{AhlswedeCsiz} and \cite{IgnaTrans} for the CS model as follows: 
\begin{alignat}{2}
M &= \Enc_2(\widetilde{X}^N,S) &&= [S^{\prime}\!+\!S, M^{\prime}] \\
\hat{S} &= \Dec_2(Y^N,M) &&= S^{\prime}\!+\!S\!-\!\hat{S^{\prime}} 
\end{alignat}
where $M'$ is the helper data for the GS model, and the addition and subtraction operations are modulo-$\displaystyle |\mathcal{S}|$. 

\emph{Error Probability}: We have
\begin{align}
\Pr[S\ne\hat{S}] = \Pr[S^{\prime}\ne\hat{S}^{\prime}] 
\end{align}
which is small by (\ref{eq:perror}).

\emph{Secrecy-leakage Rate}: The helper data $M$ of the CS model consists of $\displaystyle S^{\prime}\!+\!S$ and the helper data $M^{\prime}$ of the GS model. Using (\ref{eq:secrecyleakageresult}), (\ref{eq:keyunifresult}), and since $S$ is independent of $\displaystyle S^{\prime}M^{\prime}\mathcal{\tilde{C}}$ and uniformly distributed, we obtain
\begin{align}
\frac{1}{N}I(S;M^{\prime},S^{\prime}\!+\!S|\mathcal{\tilde{C}})\le 2(\delta_1\!-\!\delta_2\!+\!\delta_{\epsilon}).
\end{align}
We thus have a secrecy-leakage rate that is small with small $\epsilon$.

\emph{Privacy-leakage Rate}: Using (\ref{eq:privacyleaklast}), we have 
\begin{align}
&\frac{1}{N}I(X^N;M^{\prime},S^{\prime}\!+\!S|\mathcal{\tilde{C}})\nonumber\\
&\!\leq\! I(U;X)\!-\!I(U;Y)\!+\!\delta^{\prime\prime}\!+\!\delta_{\epsilon}^{\prime}\!+\!\delta_1
\end{align}
since $\displaystyle S^{\prime}\!+\!S$ is independent of $\displaystyle M^{\prime}X^N\mathcal{\tilde{C}}$.  

\emph{Storage Rate}: We obtain
\begin{align}
&\frac{1}{N}H(M^{\prime},S^{\prime}\!+\!S|\mathcal{\tilde{C}})\stackrel{(a)}{\leq}R_m+\frac{1}{N}H(S^{\prime}\!+\!S)\nonumber\\
&\stackrel{(b)}{=}I(U;\widetilde{X})\!-\!I(U;Y)\!+\!\delta_1+\!I(U;Y)\!-\!\delta_2\nonumber\\
&=I(U;\widetilde{X})+\delta_1\!-\!\delta_2
\end{align}
where $(a)$ follows because $\displaystyle S^{\prime}\!+\!S$ is independent of $\displaystyle M^{\prime}\mathcal{\tilde{C}}$ and $(b)$ follows by (\ref{eq:r_mchoose}) and (\ref{eq:r_schoose}). 

Using the selection lemma \cite[Lemma 2.2]{Blochbook}, there exists a codebook for the CS model that approaches the key-leakage-storage triple
\begin{align*}
(R_s,R_l,R_m)=\big(&I(U;Y),\nonumber\\
&I(U;X)\!-\!I(U;Y),\nonumber\\
&I(U;\widetilde{X})\big).
\end{align*}

\section{Converses}\label{sec:convProofs}
The converses for Theorems~\ref{theo:regiongenerated} and ~\ref{theo:regionchosen} follow similar steps. Therefore, we give the proofs of both theorems with different bounds for the storage rates.

Suppose that for some $\delta\!>\!0$ and $N$ there is an encoder and a decoder such that (\ref{eq:reliabilityconst})-(\ref{eq:storageconst}) are satisfied for the GS or CS model by the key-leakage-storage triple $\displaystyle (R_s,R_l,R_m)$. Fano's inequality for $S$ and $\hat{S}$ gives

\begin{align}
N\epsilon_N\!\geq\!H(S|\hat{S})\!\overset{(a)}{\geq}\!H(S|MY^N)\label{eq:fano}
\end{align}
where $\epsilon_N\!=\!\delta R_s\!+\!H_b(\delta)/N$ and $(a)$ permits randomized decoding. Note that $\epsilon_N\!\rightarrow\!0$ if $\delta\!\rightarrow\!0$. We use (\ref{eq:fano}) to bound the key, leakage, and storage rates.

\emph{Secret-key rate}: Using (\ref{eq:secrecyconst}), (\ref{eq:uniformityconst}), (\ref{eq:fano}), and because $\displaystyle Y^{i-1}\!-\!MSX^{i-1}\!-\!Y_i$ forms a Markov chain, we obtain
\begin{align}
N&(R_s\!-\!\delta)\!\leq H(S)=I(S;M)+I(S;Y^N|M)+H(S|MY^N)\nonumber\\
\leq&NH(Y)\!-\!\sum_{i=1}^{N}H(Y_i|MSX^{i-1}\!)\!+\!N(\delta\!+\!\epsilon_N\!)\label{eq:convunif1}.
\end{align}

Identify $\displaystyle U_i\!=\!MSX^{i-1}$ in (\ref{eq:convunif1}) so that $\displaystyle U_i\!-\!\widetilde{X}_i\!-\!X_i\!-\!Y_i$ forms a Markov chain, which follows since $\displaystyle P_{Y|X}$ and $\displaystyle P_{\widetilde{X}|X}$ are memoryless channels. Introduce a time-sharing random variable $\displaystyle Q\!\sim\! \text{Unif}[1\!:\!N]$ independent of other random variables. Define $X\!=\!X_Q$, $\displaystyle \widetilde{X}\!=\!\widetilde{X}_Q$, $\displaystyle Y\!=\!Y_Q$, and $U\!=\!(U_Q,\!Q)$ so that $\displaystyle U\!-\!\widetilde{X}\!-\!X\!-\!Y$ forms a Markov chain. Using (\ref{eq:convunif1}), we obtain
\begin{align}
&R_s\!\leq\! H(Y)\!-\!\frac{1}{N}\sum_{i=1}^{N}H(Y_i|U_i)\!+\!2\delta\!+\!\epsilon_N\nonumber\\
&=H(Y)\!-\!H(Y_Q|U_QQ)\!+\!2\delta\!+\!\epsilon_N\nonumber\\
&=I(U;Y)\!+\!2\delta\!+\!\epsilon_N.\label{eq:keyratetheo1}
\end{align}

\emph{Storage rate}: For the GS model, we have
\begin{align}
&N(R_m\!+\!\delta)\stackrel{(a)}{\geq}H(M)\geq H(M|Y^N)\nonumber\\
&\stackrel{(b)}{\geq} H(MSY^N)\!-\!H(Y^N)\!-\!H(S|MY^N)- H(MS|\widetilde{X}^N)\nonumber\\
&\stackrel{(c)}{\geq}\!\Big[\sum_{i\!=\!1}^{N}I(MS\widetilde{X}^{i-1};\widetilde{X}_i\!)\!-\!I(MSY^{i-1};Y_i\!)\Big]\!-\!N\epsilon_N\nonumber\\
&\stackrel{(d)}{\geq}\Big[\sum_{i\!=\!1}^{N}\!I(MSX^{i-1};\widetilde{X}_i\!)\!-\!I(MSX^{i-1};Y_i\!)\Big]\!-\!N\epsilon_N\nonumber\\
&=\Big[\sum_{i\!=\!1}^{N}\!I(U_i;\widetilde{X}_i\!)\!-\!I(U_i;Y_i)\!\Big]\!-\!N\epsilon_N \label{eq:conversegsstoragerate}
\end{align}
where $(a)$ follows by (\ref{eq:storageconst}), $(b)$ follows from the encoding step, $(c)$ follows by (\ref{eq:fano}) and because $\widetilde{X}^N$ and $Y^N$ are i.i.d., and $(d)$ follows by the Markov chains
\begin{subequations}
	\begin{align}
	&Y^{i-1}\!-\!MSX^{i-1}\!-\!Y_i\label{eq:firstMarkovchain}\\
	&X^{i-1}\!-\!MS\widetilde{X}^{i-1}\!-\!\widetilde{X}_i\label{eq:secondMarkovchain}.
	\end{align}
\end{subequations}
Using the definition of $U$ above, we obtain for the GS model
\begin{align}
R_m\geq I(U;\widetilde{X})-I(U;Y)-(\delta\!+\!\epsilon_N). \label{eq:storagetheo1}
\end{align}

For the CS model, we have
\begin{align}
&N(R_m\!+\!\delta)\!\stackrel{(a)}{\geq}\! H(M)\nonumber\\
&=I(MS;\widetilde{X}^N)\!-\!H(S|M)\!+\!H(MS|\widetilde{X}^N)\nonumber\\
&\stackrel{(b)}{\geq}I(MS;\widetilde{X}^N)\!+\!I(S;M)\nonumber\\
&\geq \sum_{i=1}^{N}I(MS\widetilde{X}^{i-1};\widetilde{X}_i)\nonumber\\
&\stackrel{(c)}{\ge}\sum_{i=1}^{N}I(MSX^{i-1};\widetilde{X}_i)\nonumber\\
&=\sum_{i=1}^{N}I(U_i;\widetilde{X}_i)\label{eq:conversecsstoragerate}
\end{align}
where $(a)$ follows by (\ref{eq:storageconst}), $(b)$ follows because S is independent of $\widetilde{X}^N$ and from the encoding step, and $(c)$ follows by applying (\ref{eq:secondMarkovchain}). Using the definition of $U$ above, we have for the CS model
\begin{align}
R_m\ge I(U;\widetilde{X})-\delta.\label{eq:storagetheo2}
\end{align}

\emph{Privacy-leakage rate}: Observe that 
\begin{align}
N&(R_l\!+\!\delta)\stackrel{(a)}{\geq}I(X^N;M)\geq H(M|Y^N)-H(M|X^N)\nonumber\\ 
=&H(MSY^N)\!-\!H(S|MY^N)\!-\!H(Y^N)\!-\!H(M|X^N)\nonumber\\
\geq&I(MS;X^N)\!-\!I(MS;Y^N)\!-\!H(S|MY^N)\nonumber\\
\stackrel{(b)}{\geq}&\Big[\sum_{i\!=\!1}^{N}I(MSX^{i-1};X_i\!)\!-\!I(MSY^{i-1};Y_i\!)\Big]\!-\!N\epsilon_N\nonumber\\
\stackrel{(c)}{\geq}&\Big[\sum_{i\!=\!1}^{N}I(MSX^{i-1};X_i\!)\!-\!I(MSX^{i-1};Y_i\!)\Big]\!-\!N\epsilon_N\nonumber\\
=&\Big[\sum_{i\!=\!1}^{N}I(U_i;X_i\!)\!-\!I(U_i;Y_i\!)\Big]\!-\!N\epsilon_N
\end{align}
where $(a)$ follows by (\ref{eq:privacyconst}), $(b)$ follows by (\ref{eq:fano}), and $(c)$ follows from the Markov chains in (\ref{eq:firstMarkovchain}). Using the definition of $U$ above, we have
\begin{align}
R_l\geq I(U;X)-I(U;Y)-(\delta\!+\!\epsilon_N)\label{eq:leakagetheo1}.
\end{align}
The converse for Theorem~\ref{theo:regiongenerated} follows by (\ref{eq:keyratetheo1}), (\ref{eq:storagetheo1}), and (\ref{eq:leakagetheo1}), and by letting $\displaystyle \delta\!\rightarrow\!0$. The converse for Theorem~\ref{theo:regionchosen} follows by (\ref{eq:keyratetheo1}), (\ref{eq:storagetheo2}), and (\ref{eq:leakagetheo1}), and by letting $\displaystyle \delta\!\rightarrow\!0$.

\section{Conclusion}\label{sec:conclusion}
We derived the key-leakage-storage regions for a HSM for noisy biometric and physical identifiers. For a BSS, we used MGL to evaluate the key-leakage-storage regions for decoder-output channels that can be decomposed into a mixture of BSCs and quantified the improvements in all rates with multiple measurements at the decoder as compared to a single measurement. By taking a large number of measurements of the hidden source at the decoder, the privacy-leakage rate is made small for the GS and CS models, and the storage rate is decreased for the GS model, which is not possible for the CS model. We showed that if one mistakenly uses the VSM when the source is hidden, the privacy-leakage rate might be pessimistic, whereas the secret-key and storage rates might be over-optimistic, which leads to unnoticed secrecy leakage and reliability reductions. The points that achieve the maximum secret-key rates in the key-leakage-storage regions for multiple encoder measurements show that the gain in the secret-key rate from multiple encoder measurements results in greater privacy-leakage and significantly greater storage rates.  

The examples illustrated that higher reliability in the encoder measurements improves the storage rate, which also applies to decoder measurements because the encoder and decoder measurements are obtained through separate channels. In future work, we plan to consider key-leakage-storage regions for encoder and decoder measurements through a broadcast channel, and to show that reduced reliability in the measurements might enlarge the key-leakage-storage region for this case.

\section*{Acknowledgment}
O. G\"unl\"u thanks Roy Timo and Kittipong Kittichokechai for useful discussions and insightful comments. 

\appendices

\section{Cardinality Bound}\label{app:cardinalitybound}
Consider $\displaystyle \mathcal{\widetilde{X}}\!=\!\{\tilde{x}_1,\tilde{x}_2,\ldots,\!\tilde{x}_{|\mathcal{\widetilde{X}}|}\}$ and the following $\displaystyle |\mathcal{\widetilde{X}}|\!+\!2$ real-valued continuous functions on the connected compact subset $\displaystyle \mathcal{P}$ of all probability distributions on $\displaystyle \mathcal{\widetilde{X}}$:
\begin{align}
 f_j(P_{\widetilde{X}})=\begin{cases}
               P_{\widetilde{X}}(\tilde{x}_j) \text{ for } j=1,2,\ldots,|\mathcal{\widetilde{X}}|\!-\!1\\
               H(X) \text{ for } j=|\mathcal{\widetilde{X}}|\\
               H(\widetilde{X}) \text{ for } j=|\mathcal{\widetilde{X}}|+1\\
               H(Y) \text{ for } j=|\mathcal{\widetilde{X}}|+2.
            \end{cases}
\end{align}

By using the support lemma \cite[Lemma 15.4]{CsiszarKornerbook2011}, we find that there is a random variable $\displaystyle U^{\prime}$ taking at most $\displaystyle |\mathcal{\widetilde{X}}|\!+\!2$ values such that $\displaystyle P_{\widetilde{X}}$, $\displaystyle H(\widetilde{X})$, $\displaystyle H(X|U)$, $\displaystyle H(\widetilde{X}|U)$, and $\displaystyle H(Y|U)$ are preserved if we replace $U$ with $\displaystyle U^{\prime}$. We preserve the joint distribution $P_{\widetilde{X}XY}(\tilde{x},x,y)\!=\!P_{\widetilde{X}}(\tilde{x})P_{X|\widetilde{X}}(x|\tilde{x})P_{Y|X}(y|x)$ by preserving $P_{\widetilde{X}}(\tilde{x})$, so the entropies $\displaystyle H(X)$ and $H(Y)$ are also preserved. Hence, the expressions in Theorems~\ref{theo:regiongenerated} and \ref{theo:regionchosen}
\begin{align}
 &I(U;Y) \!=\! H(Y)\!-\!H(Y|U)\nonumber\\
 &I(U;X)-I(U;Y) \!=\! H(X)\!-\!H(X|U)\!-\!H(Y)\!+\!H(Y|U)\nonumber\\
 & I(U;\widetilde{X})\!-\!I(U;Y) \!=\!H(\widetilde{X})\!-\!H(\widetilde{X}|U)\!-\! H(Y)\!+\!H(Y|U)\nonumber\\
 &I(U;\widetilde{X}) \!=\!H(\widetilde{X})\!-\!H(\widetilde{X}|U)\nonumber
\end{align}
are preserved by some $\displaystyle U^{\prime}$ that satisfies the Markov condition $U\!-\!\widetilde{X}\!-\!X\!-\!Y$ with $\displaystyle |\mathcal{U^{\prime}}|\leq|\mathcal{\widetilde{X}}|+2$. 

\section{On A Lower Bound for Binary Asymmetric Channels}\label{app:nottight}
Consider a Markov chain $U\!-\!X\!-\!Y_1$, a binary random variable $X$ with the probability distribution $Q_X$, and a binary channel $P_{Y_1|X}$ with probability transition matrix
\begin{align}
 T = \begin{bmatrix}
       a & 1-a          \\[0.3em]
       b & 1-b
     \end{bmatrix}\label{eq:transmatrix}
\end{align}
which is asymmetric if $a\!+\!b\!\ne\!1$. One can restrict attention to the cases $a\!+\!b\!\leq\!1$ and $a\!\geq\! b$ by swapping the outputs and inputs, respectively, if necessary \cite{WitzenEIfDC}. The conditional entropies $H(X|U)$ and $H(Y_1|U)$ are as defined in (\ref{eq:hxgivenu}) and (\ref{eq:hygivenu}), respectively. Since the convexity property is satisfied for all binary channels $P_{Y_1|X}$ (see Section~\ref{sec:multimeasurecap}), we have the following lower bound due to Lemma~\ref{lem:lemmaconvex}: 
\begin{align}
 &H(Y_1|U) \nonumber\\
 &\geq\! H_b\bigg(aH_b^{-1}(H(X|U))\!+\!b\big(1\!-\!H_b^{-1}(H(X|U))\big)\bigg).\label{eq:convexboundforbinary}
\end{align}
Note that if $a\!=\!b$, then (\ref{eq:convexboundforbinary}) does not depend on $H(X|U)$, since the channel would then have zero capacity.

Consider the achievability of the bound in (\ref{eq:convexboundforbinary}) for the model considered in \cite[Theorem 1]{bizimCNS}, where a ternary $U$ suffices to evaluate the key-leakage-storage region. For a channel $P_{X|U}$ with probability transition matrix 
\begin{align}
  T_u = \begin{bmatrix}
       a_u & 1-a_u          \\[0.3em]
       b_u & 1-b_u          \\[0.3em]
       c_u & 1- c_u
     \end{bmatrix}
\end{align}
and $i\!\in\!\{1,2,\ldots,|\mathcal{U}|\}$, we obtain 
\begin{align}
 &H(X|U) =  P_U(u_0)H_b(a_u)\!+\!P_U(u_1)H_b(b_u)\nonumber\\
 &\qquad\quad\qquad\!+\!P_U(u_2)H_b(c_u)\\
 &H(Y_1|U) =    P_U(u_0) H_b(aa_u\!+\!b(1\!-\!a_u))\nonumber\\
 &\qquad\qquad\quad +P_U(u_1) H_b(ab_u\!+\!b(1\!-\!b_u))\nonumber\\
 &\qquad\qquad\quad +P_U(u_2) H_b(ac_u\!+\!b(1\!-\!c_u))
\end{align}
where $P_U(u_2)\!=\!1\!-\!P_U(u_0)\!-\!P_U(u_1)$ and 
\begin{align}
 P_U(u_1)\! =\! \frac{P_X(0)\!-\!c_u\!-\!P_U(u_0)(a_u\!-\!c_u)}{b_u\!-\!c_u}.
\end{align}

\begin{figure}[t]
\centering
 \includegraphics[width=0.54\textwidth, height=0.671\textheight,keepaspectratio]{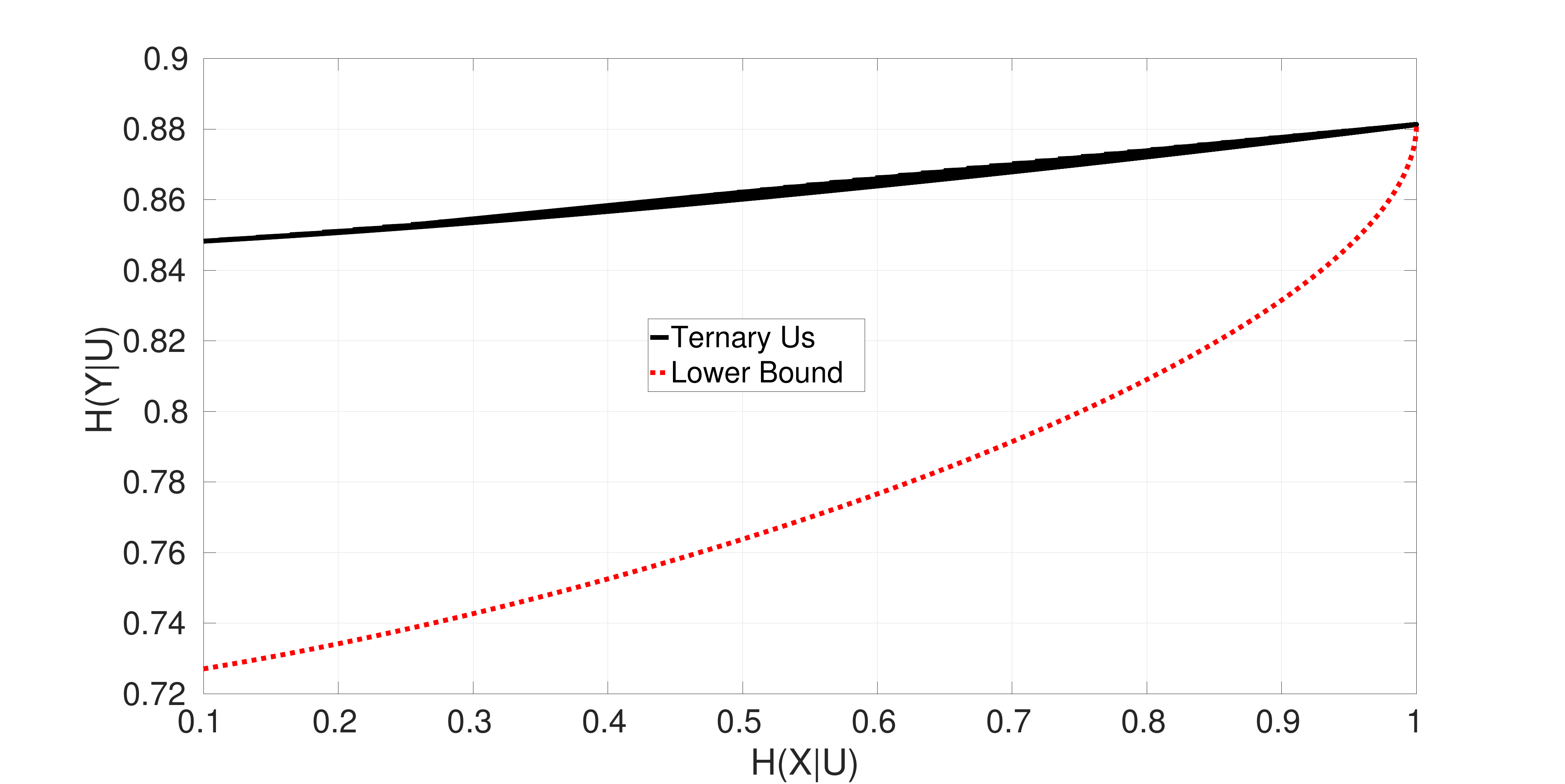}
\caption{Comparison of the lower bound and possible choices of $U$ for a uniform input and a binary channel $P_{Y_1|X}$ with parameters $a\!=\!0.4$ and $b\!=\!0.2$ in (\ref{eq:transmatrix}).} \label{fig:binarynottight} 
\end{figure} 

Fig.~\ref{fig:binarynottight} shows the possible ($H(X|U)$, $H(Y_1|U)$) pairs by assigning an appropriate set of values to the first column of $T_u$ and to $P_U(u_0)$ for a uniform $X$ and an asymmetric binary channel $P_{Y_1|X}$ with parameters $a\!=\!0.4$ and $b\!=\!0.2$. The lower bound in Fig.~\ref{fig:binarynottight} is thus not tight for such an asymmetric binary channel. Our simulations suggest that this is the case for all asymmetric channels, except for special cases like $a\!=\!b$.

\ifCLASSOPTIONcaptionsoff
  \newpage
\fi

\bibliographystyle{IEEEtran}
\bibliography{IEEEabrv,references}

\vspace*{1cm}
\begin{IEEEbiography}[{\includegraphics[width=1.05in,height=1.22in]{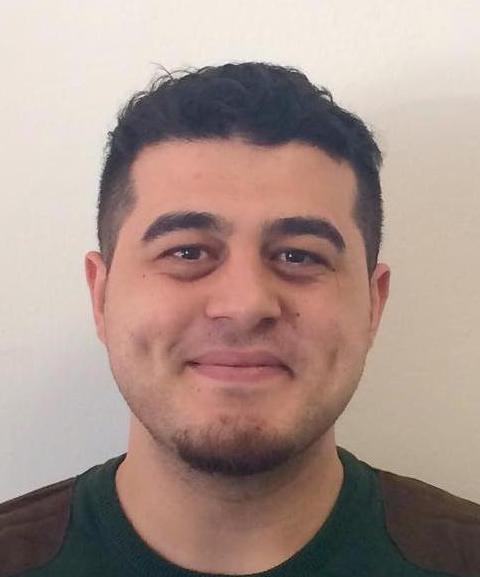}}]{Onur G\"unl\"u}
	(S'10) received the B.Sc. degree in electrical and electronics engineering from Bilkent University, Ankara, in 2011, and the M.Sc. degree in communications engineering from the Technical University of Munich (TUM), Munich, in 2013, where he is currently pursuing the Dr.-Ing. degree. He is a Research and Teaching Assistant with TUM Chair of Communications Engineering. In 2018, he visited the Information and Communication Theory Lab, TU Eindhoven, The Netherlands. His research interests include information theoretic privacy and security, code design for key agreement, statistical signal processing for biometric secrecy systems and physical unclonable functions (PUFs).
\end{IEEEbiography}

\begin{IEEEbiographynophoto}{Gerhard Kramer}
	(S'91\textendash M'94\textendash SM'08\textendash F'10) received the Dr. sc. techn. degree from ETH Zurich in 1998. From 1998 to 2000, he was with Endora Tech AG in Basel, Switzerland, and from 2000 to 2008 he was with the Math Center at Bell Labs in Murray Hill, NJ, USA. He joined the University of Southern California, Los Angeles, CA, USA, as a Professor of Electrical Engineering in 2009. He joined the Technical University of Munich (TUM) in 2010, where he is currently Alexander von Humboldt Professor and Chair of Communications Engineering. His research interests include information theory and communications theory, with applications to wireless, copper, and optical fiber networks. Dr. Kramer served as the 2013 President of the IEEE Information Theory Society.
\end{IEEEbiographynophoto}

\end{document}